\documentclass[twocolumn]{aastex62}

\begin{document}

\title{Methodology and Performance of the Two-Year Galactic Plane Scanning Survey of \emph{Insight-HXMT}}

\author{Na Sai}
\altaffiliation{Corresponding author \\Email addresses: saina@ihep.ac.cn, liaojinyuan@ihep.ac.cn}
\affiliation{Key Laboratory of Particle Astrophysics, Institute of High Energy Physics, Chinese Academy of Sciences, 19B Yuquan Road, Beijing 100049, People’s Republic of China}
\affiliation{University of Chinese Academy of Sciences, Chinese Academy of Sciences, Beijing 100049, People’s Republic of China}

\author{Jin-Yuan Liao}
\altaffiliation{Corresponding author \\Email addresses: saina@ihep.ac.cn, liaojinyuan@ihep.ac.cn}
\affiliation{Key Laboratory of Particle Astrophysics, Institute of High Energy Physics, Chinese Academy of Sciences, 19B Yuquan Road, Beijing 100049, People’s Republic of China}

\author{Cheng-Kui Li}
\affiliation{Key Laboratory of Particle Astrophysics, Institute of High Energy Physics, Chinese Academy of Sciences, 19B Yuquan Road, Beijing 100049, People’s Republic of China}

\author{Ju Guan}
\affiliation{Key Laboratory of Particle Astrophysics, Institute of High Energy Physics, Chinese Academy of Sciences, 19B Yuquan Road, Beijing 100049, People’s Republic of China}

\author{Chen Wang}
\affiliation{Key Laboratory of Space Astronomy and Technology, National Astronomical Observatories, Chinese Academy of Sciences, Beijing 100012, People’s Republic of China}
\affiliation{University of Chinese Academy of Sciences, Chinese Academy of Sciences, Beijing 100049, People’s Republic of China}

\author{Yi Nang}
\affiliation{Key Laboratory of Particle Astrophysics, Institute of High Energy Physics, Chinese Academy of Sciences, 19B Yuquan Road, Beijing 100049, People’s Republic of China}
\affiliation{University of Chinese Academy of Sciences, Chinese Academy of Sciences, Beijing 100049, People’s Republic of China}

\author{Yuan Liu} 
\affiliation{Key Laboratory of Space Astronomy and Technology, National Astronomical Observatories, Chinese Academy of Sciences, Beijing 100012, People’s Republic of China}

\author{Cheng-Cheng Guo} 
\affiliation{Key Laboratory of Particle Astrophysics, Institute of High Energy Physics, Chinese Academy of Sciences, 19B Yuquan Road, Beijing 100049, People’s Republic of China}
\affiliation{University of Chinese Academy of Sciences, Chinese Academy of Sciences, Beijing 100049, People’s Republic of China}

\author{Shu Zhang}
\affiliation{Key Laboratory of Particle Astrophysics, Institute of High Energy Physics, Chinese Academy of Sciences, 19B Yuquan Road, Beijing 100049, People’s Republic of China}

\author{Shuang-Nan Zhang}
\affiliation{Key Laboratory of Particle Astrophysics, Institute of High Energy Physics, Chinese Academy of Sciences, 19B Yuquan Road, Beijing 100049, People’s Republic of China}
\affiliation{University of Chinese Academy of Sciences, Chinese Academy of Sciences, Beijing 100049, People’s Republic of China}
\affiliation{Key Laboratory of Space Astronomy and Technology, National Astronomical Observatories, Chinese Academy of Sciences, Beijing 100012, People’s Republic of China}






\begin{abstract}

The Galactic plane scanning survey is one of the main scientific objectives of \emph{the Hard X-ray Modulation Telescope} (known as \emph{Insight-HXMT}). During the two-year operation of \emph{Insight-HXMT}, more than 1000 scanning observations have been performed and the whole Galactic plane ($\rm 0^{\circ}<l<360^{\circ}$, $\rm -10^{\circ}<b<10^{\circ}$) has been covered completely. We summarize the Galactic plane scanning survey of \emph{Insight-HXMT} for two years, including the characteristics of the scanning data, the data analysis process and the preliminary results of the Low-Energy telescope, the Medium-Energy telescope and the High-Energy telescope. With the light curve PSF fitting method, the fluxes of the known sources in the scanned area as well as the flux errors are obtained for each scanning observation. From the relationships of SNRs and fluxes, the $5\sigma$ sensitivities of three telescopes of \emph{Insight-HXMT} are estimated as $\rm \sim7.6\times10^{-11}~erg~cm^{-2}~s^{-1}$ ($\rm 3~mCrab,~1-6~keV$), $\rm \sim4.0\times10^{-10}~erg~cm^{-2}~s^{-1}$ ($\rm 20~mCrab,~7-40~keV$) and $\rm \sim2.6\times10^{-10}~erg~cm^{-2}~s^{-1}$ ($\rm 18~mCrab,~25-100~keV$) for an individual scanning observation of $2-3$ hours, respectively. Up to September 2019, more than 800 X-ray sources with various types are monitored by the three telescopes and their long-term light curves with three energy bands are obtained to make further scientific analyses. 

\end{abstract}



\section{Introduction} \label{sec:intro}
\emph{The Hard X-ray Modulation Telescope} (\emph{Insight-HXMT}) is China's first space astronomical satellite launched on June 15th, 2017. The Galactic plane scanning survey is one of the main scientific objectives of \emph{Insight-HXMT} to search for new transients and monitor the known variable sources in a wide X-ray energy band. From the previous surveys by other X-ray telescopes, e.g., \emph{ROSAT}, \emph{INTEGRAL} and \emph{Swift}, most of the hard X-ray radiation objects in the Galactic plane are variable sources, and mainly X-ray binaries \citep{2013ApJS...207...19, 2016ApJS...223...15}. Thanks to the design of its large effective area and narrow field of view in X-ray band, \emph{Insight-HXMT} has advantages in the weak and variable source survey.

\begin{figure*}[htbp]
\centering
\gridline{\fig{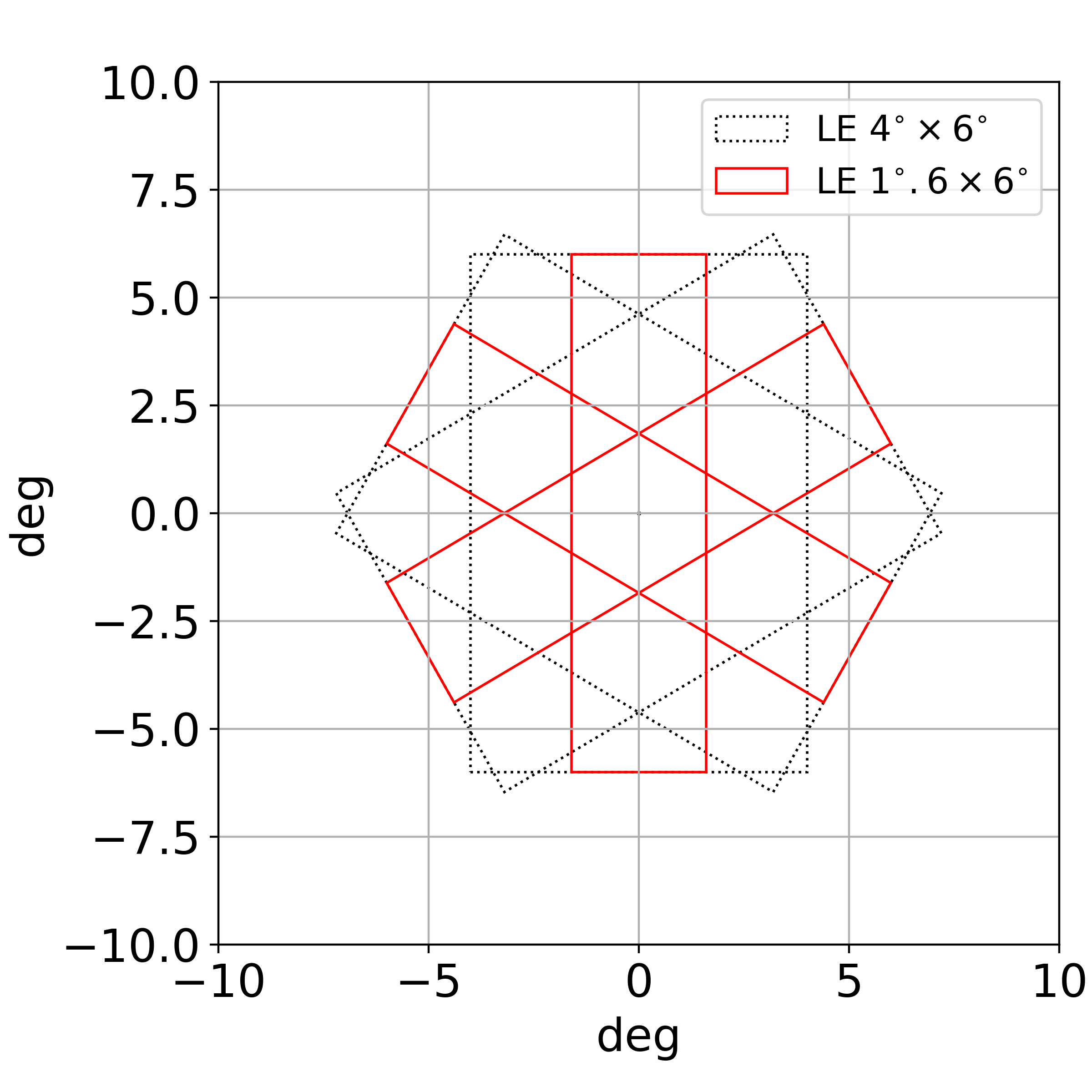}{0.3\textwidth}{(a)}
\fig{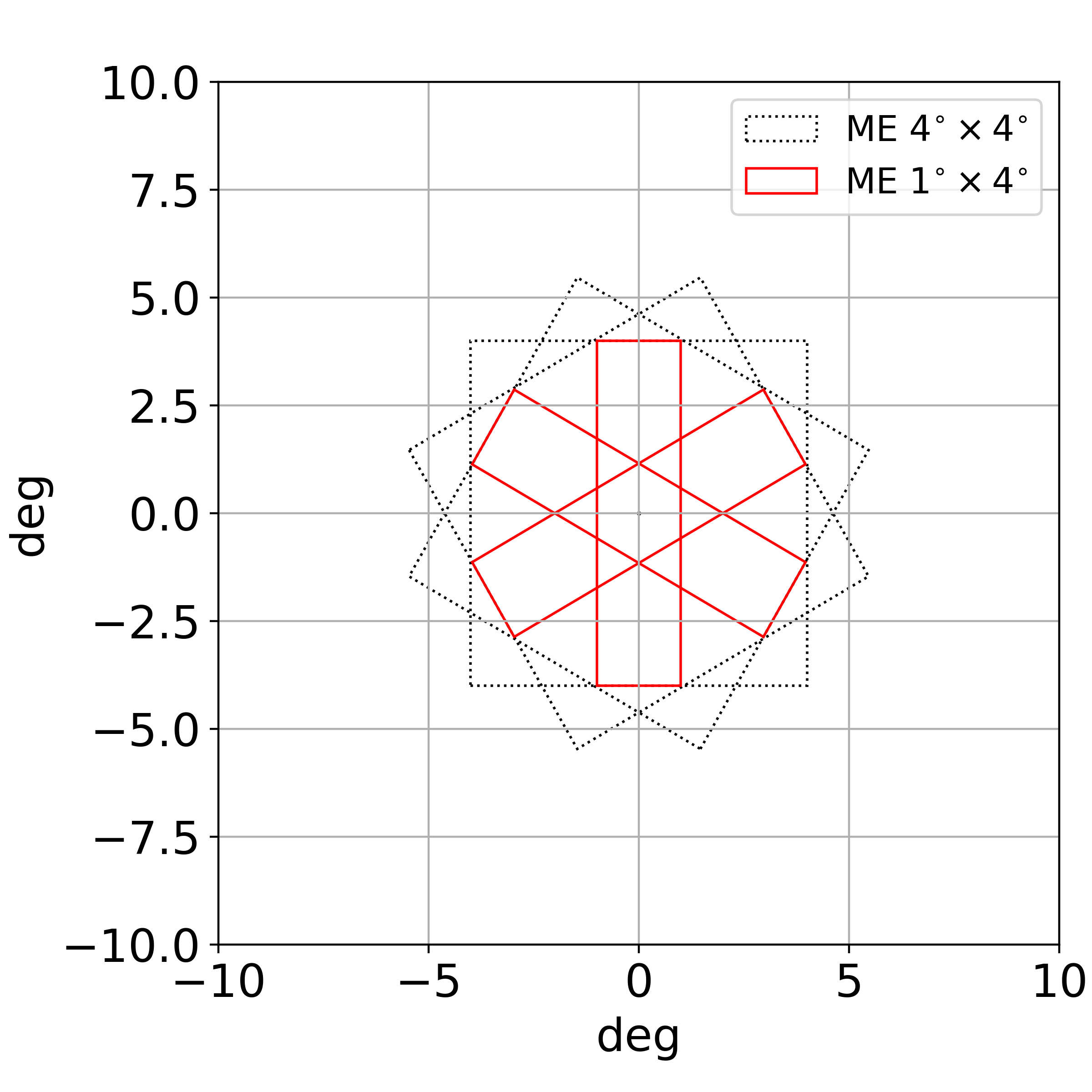}{0.3\textwidth}{(b)}
\fig{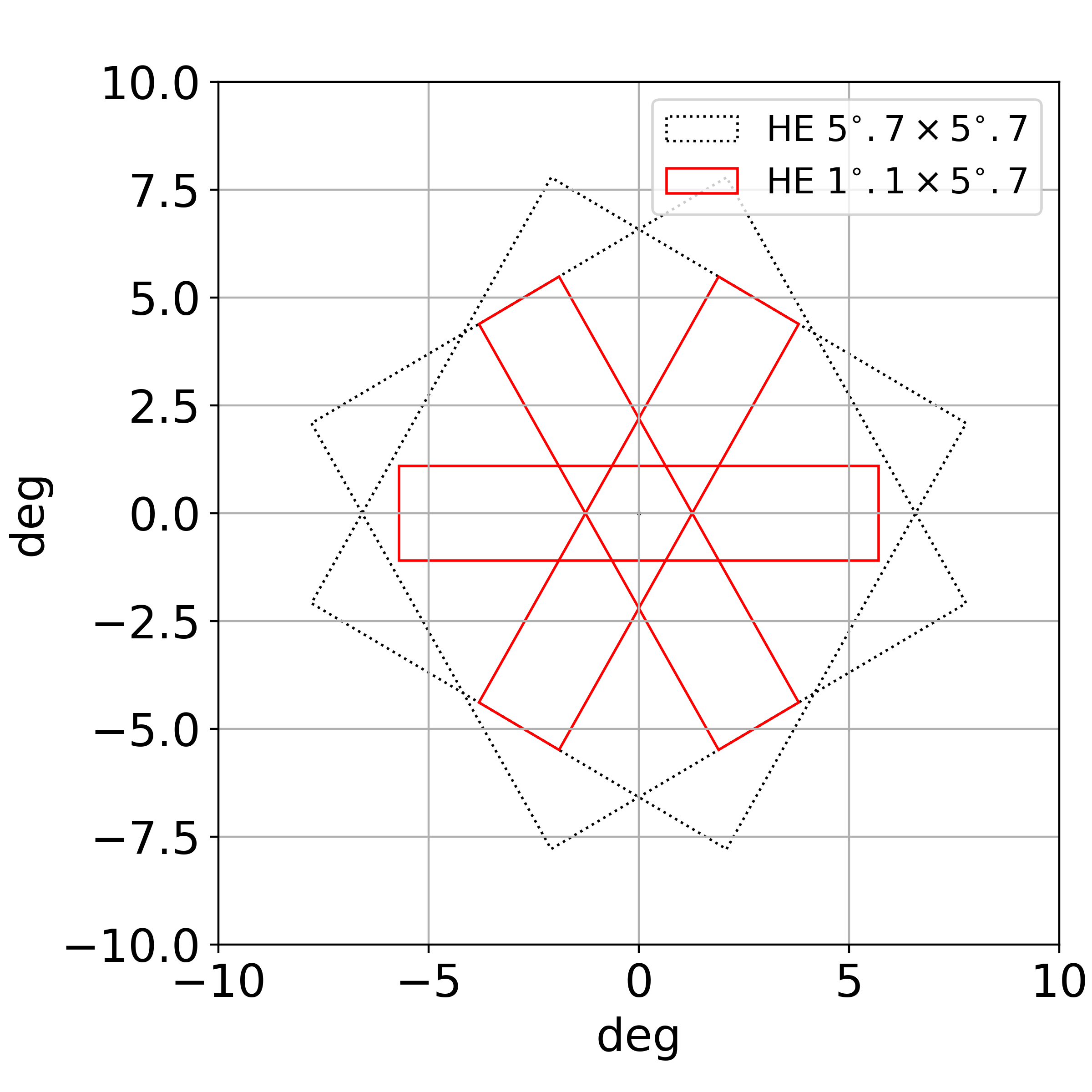}{0.3\textwidth}{(c)}}
\caption{FOVs of LE, ME and HE of \emph{Insight-HXMT}. \label{fig:fov}}
\end{figure*}

There are three main payloads onboard \emph{Insight-HXMT}, i.e. the Low-Energy X-ray telescope (LE, $0.7-13$~keV), the Medium-Energy X-ray telescope (ME, $5-40$~keV) and the High-Energy X-ray telescope (HE, $20-250$~keV) \citep{2019Zhang}. LE is composed of three detector boxes, and each includes 8 detector modules \citep{2019Chen}. ME is composed of three detector boxes with the Si-PIN detector arrays \citep{2019Cao}. HE consists of 18 NaI(Tl)/CsI(Na) phoswich detectors \citep{2019Liu}. All the instruments are collimated telescopes and mainly composed of the detectors with three small filed of views (FOVs) differing by $60^{\circ}$. There are one box of LE, one box of ME and five detectors of HE in each small FOV. In addition, the large FOV detectors also operate as supplements (Figure \ref{fig:fov}).
Thanks to the relatively narrow FOV, the small FOV detectors can obtain more accurate source fluxes and positions than the large FOV detectors in the scanning observations.
We only use the small FOV detectors in the Galactic plane scanning survey.

\begin{deluxetable}{cccc}[htbp]
\tablecaption{Main parameters of \emph{Insight-HXMT} \label{tab:payload}}
\tablecolumns{5}
\tablenum{1}
\tablewidth{0pt}
\tablehead{
\colhead{ } &
\colhead{LE} &
\colhead{ME} &
\colhead{HE}
}
\startdata
Geometrical area ($\rm cm^2$) & $384$ & $952$ & $5096$ \\
FOV\tablenotemark{a} (FWHM) & $1^{\circ}.6\times6^{\circ}$ & $1^{\circ}\times4^{\circ}$ & $1^{\circ}.1\times5^{\circ}.7$ \\
Energy range (keV) & $0.7-13$ & $5-40$ & $20-250$ \\
Energy range\tablenotemark{b} (keV) & $1-6$ & $7-40$ & $25-100$ \\
\enddata
\tablenotetext{a}{Small FOV.}
\tablenotetext{b}{Used in the Galactic plane scanning survey.}
\end{deluxetable}

The main objectives of the \emph{Insight-HXMT} Galactic plane scanning survey are: 
\begin{enumerate}
\item Monitoring the known sources in the Galactic plane. 
\item Searching for new X-ray transients.
\item Studying the activity mechanism of the compact objects and exploring the distribution and formation in the Galaxy. 
\end{enumerate}

\begin{figure}[htbp]
\centering
\plotone{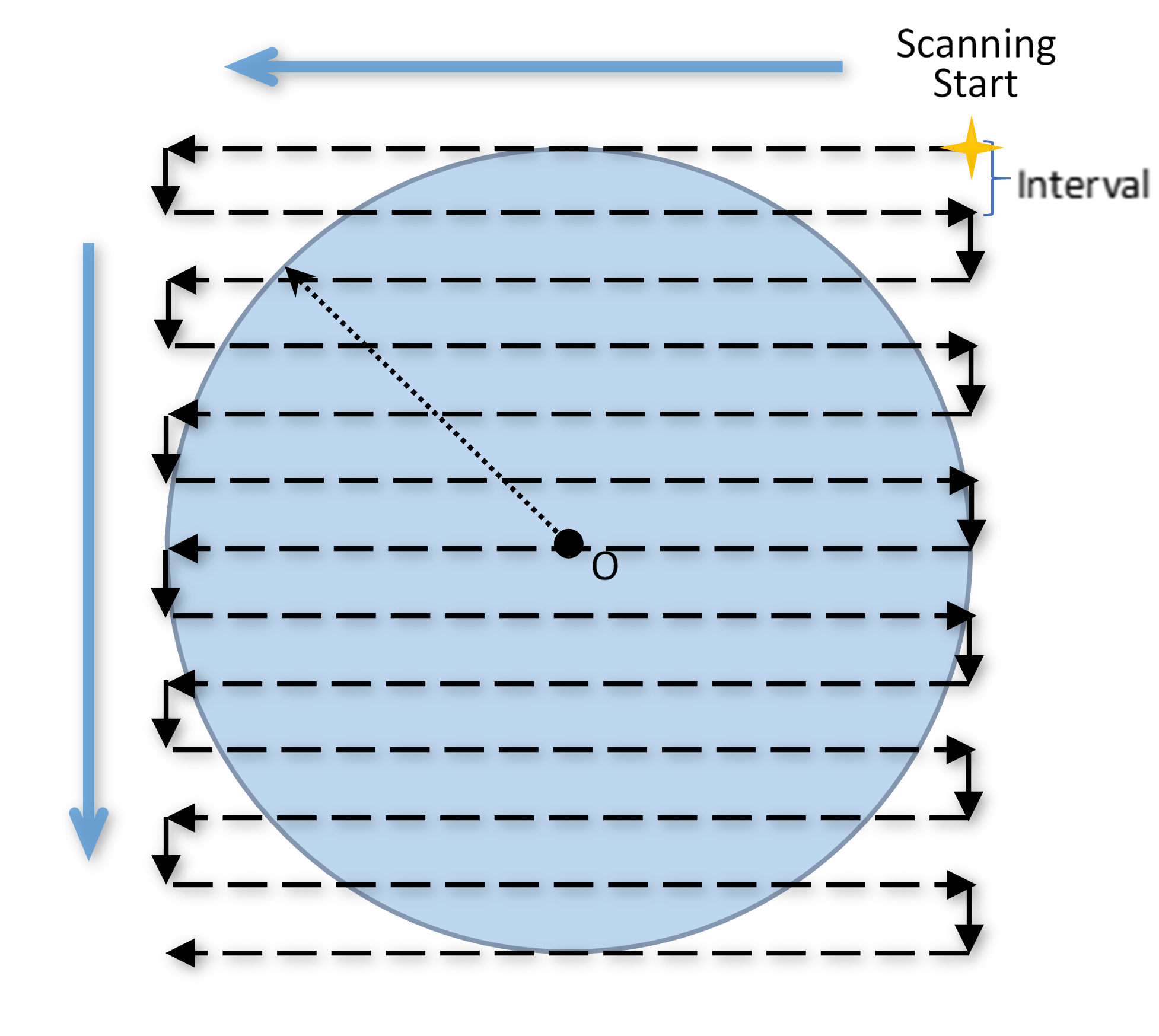}
\caption{Schematic diagram of a small area scan. ``O'' is the center of the scanned area and the circle is the observed area. The dotted line with the arrow is the scanning track.\label{fig:scan}}
\end{figure}

In the Galactic plane scanning survey of \emph{Insight-HXMT}, the whole Galactic plane ($\rm 0^{\circ}<l<360^{\circ}$, $\rm -10^{\circ}<b<10^{\circ}$) is divided into several areas of the same radius. In order to cover the whole Galactic plane completely without gaps, there is an overlap between the adjacent areas. Due to the limit of the solar angle ($\rm \emph{A}_s>70^{\circ}$), the visible time of each area is about half a year. For every scanned area, the progressive scanning mode (Figure \ref{fig:scan}) is used to perform the scanning observation. There are three scanning speeds ($\rm \emph{v}=0^{\circ}.01~s^{-1},~0^{\circ}.03~s^{-1},~0^{\circ}.06~s^{-1}$) and 10 scanning intervals ($\rm \emph{d}=0^{\circ}.1,~0^{\circ}.2~…~to~1^{\circ}$) that can be used in scanning observations. The duration of an individual scanning observation is generally $2-3$~hours, which depends on the scanning parameters. For a scanning observation with $\emph{R}=10^{\circ}$, $\rm \emph{v}=0^{\circ}.06~s^{-1}$, and $\emph{d}=0^{\circ}.8$, the duration is $\sim3.3$~hours.

In this paper, we summarize the two-year Galactic plane scanning survey of \emph{Insight-HXMT} as follows. In Section \ref{sec:data process}, we present the details of data reduction. The light curve fitting is described in Section \ref{sec:lc analyse}. The preliminary results and conclusion are given in Sections \ref{sec:result} and \ref{sec:con}, respectively.

\section{Data Reduction} \label{sec:data process}
\subsection{Preliminary reduction and basic parameter selection} \label{subsec:basic process}
\begin{figure}[htbp]
\centering
\includegraphics[width=8.5 cm]{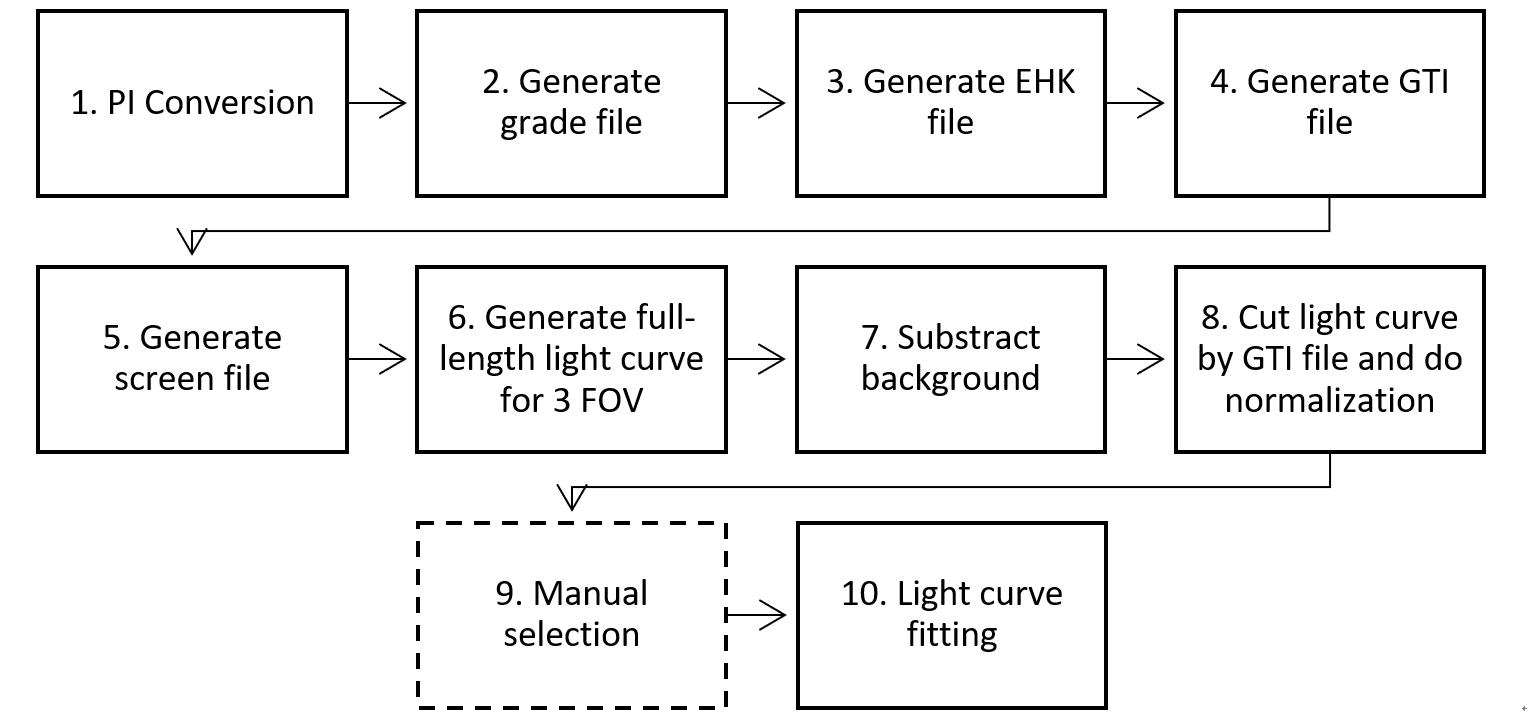}
\caption{A complete analysis process of the scan data. The artificial selection in step 9 is only need in some LE data reduction as sometimes the LE data is very complicated.\label{fig:process}}
\end{figure}

The \emph{Insight-HXMT} data analysis software HXMTDAS v2.0 \citep{2019...2.01} is used to do the preliminary data reduction. The process is the steps $1-6$ shown in the flow chart (Figure \ref{fig:process}). The Statistics-sensitive Non-linear Iterative Peak-clipping (SNIP) method \citep{1988PRB...34...396} is used to do the background subtraction for LE since the background intensity is lower. The backgrounds of HE and ME are modulated by geomagnetic field significantly \citep{2009CAA...33...333, 2015ASS...360...13} and can be fitted by a polynomial. The net light curves obtained after step 8 or 9 are used to monitor the known sources and search new X-ray transients. The event files, temperature files, voltage files and EHK (Extended housekeeping) files are used to produce the light curves with the standard process of HXMTDAS. The basic parameters used by HXMTDAS in the Galactic plane scanning survey can be divided into two parts. One is for the energy band selection of a light curve (Table \ref{tab:payload}), and another is for the good time interval (GTI) selection (Table \ref{tab:gtipar}). 

\begin{deluxetable}{cccc}[htbp]
\tablecaption{Parameters in GTI selection \label{tab:gtipar}}
\tablecolumns{9}
\tablenum{2}
\tablewidth{0pt}
\tablehead{
\colhead{ } &
\colhead{LE} &
\colhead{ME} &
\colhead{HE}
}
\startdata
ELV & $>~7^{\circ}$ & $>~5^{\circ}$ & $>~6^{\circ}$ \\
DYE\underline{ }ELV & $>~10^{\circ}$\tablenotemark{*} & $>~5^{\circ}$ & $>~6^{\circ}$ \\
 & $>~20^{\circ}$ & & \\
T\underline{ }SAA & $>100$~s & $\geqslant 200$~s & \nodata \\
TN\underline{ }SAA & $> 100$~s & $\geqslant 100$~s & \nodata \\
COR & $> 3 $~GV & $> 8 $~GV & \nodata \\
FrocedEVT & $\rm \geqslant 3000~cts~s^{-1}$ & \nodata & \nodata \\
\enddata
\tablecomments{ELV: earth elevation of FOV center position; DYE\underline{ }ELV: day earth elevation of FOV center position; T\underline{ }SAA: time since the last passage from SAA; TN\underline{ }SAA: Time to next SAA passage; COR: geomagnetic cut-off rigidity; FrocedEVT: count rate of the forced trigger event, only for LE. ``...'' denote the parameters that are not used in GTI selection.}
\tablenotetext{*}{Before 2019-09-01.}
\end{deluxetable}

The earth elevations (ELV) are selected as $7^{\circ}$, $5^{\circ}$ and $6^{\circ}$ for LE, ME and HE, respectively, to avoid the earth occultation of the scan area. The detectors will be saturated as a large number of optical photons enter the collimator, if the angle between the FOV and the bright earth (DYE\underline{ }ELV) is too small. Thus we choose $\rm DYE\underline{~}ELV=10^{\circ}$, $5^{\circ}$, and $6^{\circ}$ for LE, ME and HE, respectively. To avoid the irradiation damage of the LE detectors, DYE\underline{ }ELV for LE is adjusted to $20^{\circ}$ from September, 2019.

The flux of the charged particles is very high in the South Atlantic Anomaly (SAA) due to the low strength of the magnetic field \citep{2006CUP} that results in the high background level. LE, ME and HE of \emph{Insight-HXMT} are shutdown for protection when the satellite passes through SAA. As the backgrounds are usually unstable for LE and ME near SAA, the time since the last passage from SAA (T\underline{ }SAA) and the time to the next SAA passage (TN\underline{ }SAA) are generally selected to be $\rm 100~s~\&~100~s$ for LE and $\rm 100~s~\&~200~s$ for ME in order to avoid the influence of SAA. The HE background level is high but stable near SAA, thus T\underline{ }SAA and TN\underline{ }SAA can be ignored for the HE GTI selection.

The geomagnetic cut-off rigidity (COR) is inversely related to the particle flux, as well as the background levels of the three telescopes. The energy range of LE is chosen between $1-6$~keV, where the particle background is not dominant. For ME, the background is very sensitive to the particle flux. Thus the $\rm COR = 3~GV$ and $8$~GV are chosen in GTI selection of LE and ME, respectively. The HE background level is high for low COR; however, the high level background cannot affect the data analysis, as the background is stable. Thus the COR is not considered in the HE GTI selection.

\begin{figure}[htbp]
\flushleft
\includegraphics[width=8.5 cm]{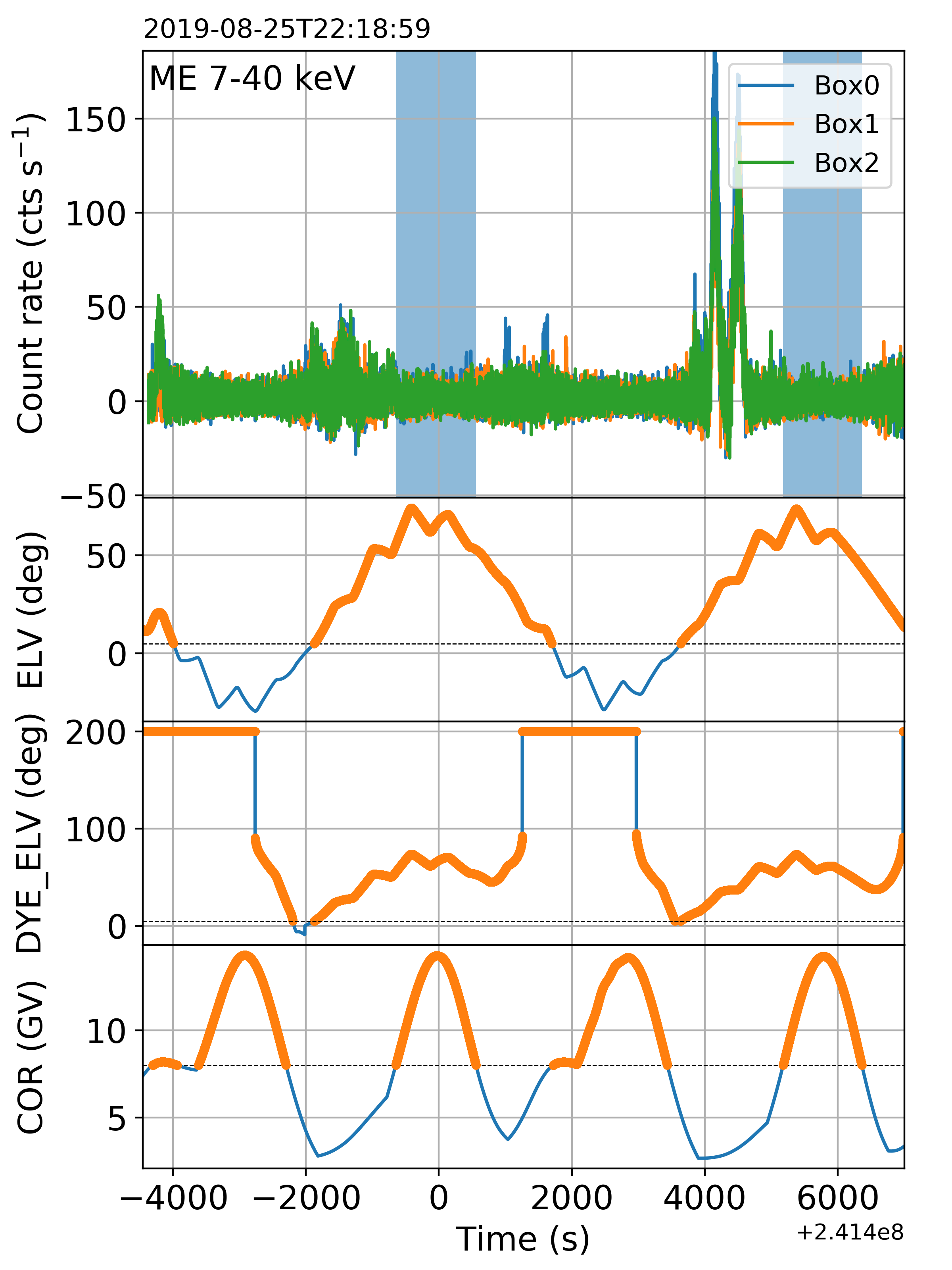}
\caption{An example of the ME preliminary reduction with the ELV, DYE\underline{ }ELV and COR. The parameter color satisfy the ME criteria in Table \ref{tab:gtipar} are shown with yellow. The preliminary GTI satisfying all the ME criteria simultaneously is marked with shadows. \label{fig:megti}}
\end{figure}

In addition, the count rate of the forced trigger event is used in the LE GTI selection and the latitude (SAT\underline{ }LAT) and longitude (SAT\underline{ }LON) of the satellite in geographical coordinate system are used in the ME GTI selection, where SAT\underline{ }LAT is from $-36^{\circ}.5$ to $36^{\circ}.5$, but excluding the rectangular area with $\rm 31^{\circ}<SAT\underline{~}LAT<38^{\circ}$ and $\rm 228^{\circ}< SAT\underline{~}LON<245^{\circ}$. All the parameters in Table \ref{tab:gtipar} are selected by the on-ground simulation and in-orbit observation after the \emph{Insight-HXMT} launch. In the selected energy bands (the last column of Table \ref{tab:payload}), there is no noise peak or other instrumental anomaly in the spectra of almost all detectors. Figure \ref{fig:megti} is an example of ME GTI preliminary selection with the main GTI parameters. The HE data can be reduced directly only with the criteria in Table \ref{tab:gtipar}, but the data of LE and ME must be further extracted.

\subsection{Special Filter of LE Data} \label{subsec:legti}
\begin{figure}[htbp]
\flushleft
\includegraphics[width=8.5 cm]{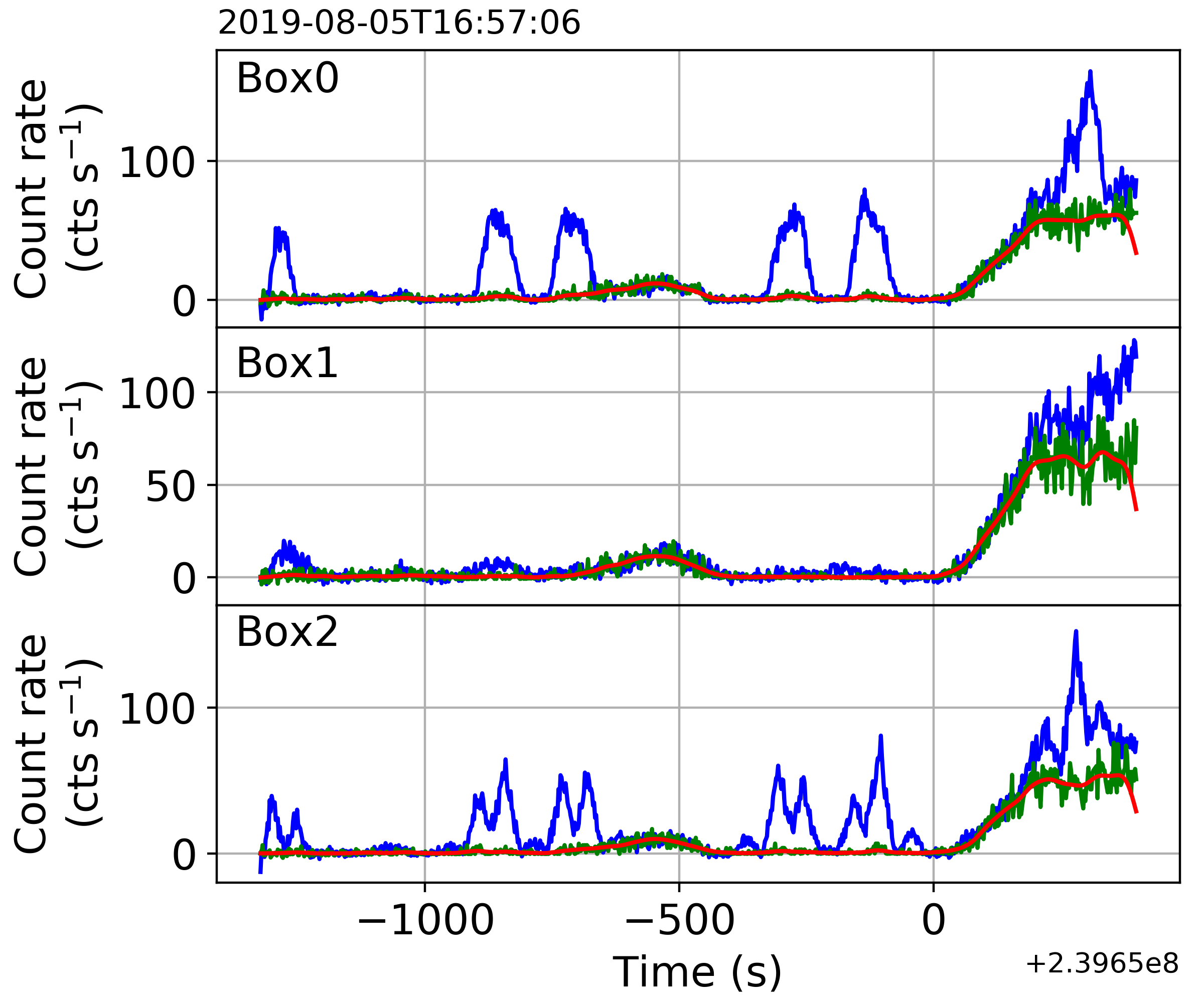}
\caption{LE light curves affected by particle events. The blue and green lines are the light curves in $1-6$~keV and $8.6-10.3$~keV, respectively. The red lines are the smoothed green lines. The green bumps on the right end are caused by particle events and need to be removed.\label{fig:lehp}}
\end{figure}

During a scanning observation, both the scanned sources and the particle events can leave peaks (bumps) in the light curves; however, the two types of peaks are very different. On one hand, the profiles of the particle peaks in the light curves with the three FOVs are very similar with no delay. On the other hand, the peaks caused by a scanned source are very different between the light curves with the three FOVs that because the times of a source passing in and out the FOVs are different, as well as the detection efficiencies when the source crosses the FOVs. Therefore, according to the simultaneity and similarity between the light curves with different FOVs, we can identify a peak caused by a particle event or by a scanned source. We found that the characteristic of the light curve caused by particle events in the high energy band is similar with that in the low energy band, which is consistent with the characteristics of the LE particle backgrounds \citep{2019Liao}. Figure \ref{fig:lehp} shows the scanning light curves with the three FOVs in $1-6$~keV and $8.6-10.3$~keV, respectively. There is no peak caused by the scanned sources in the light curves in $8.6-10.3$~keV due to the fewer source photons and the lower effective area in this energy band. The small and big bumps in the two energy bands are very similar that means they are both caused by particle events. Because the background caused by particle events is high and difficult to be estimated accurately, the part of the light curve affected by the particle events is removed from the LE GTI.

\begin{figure}[htbp]
\flushleft
\includegraphics[width=8.5 cm]{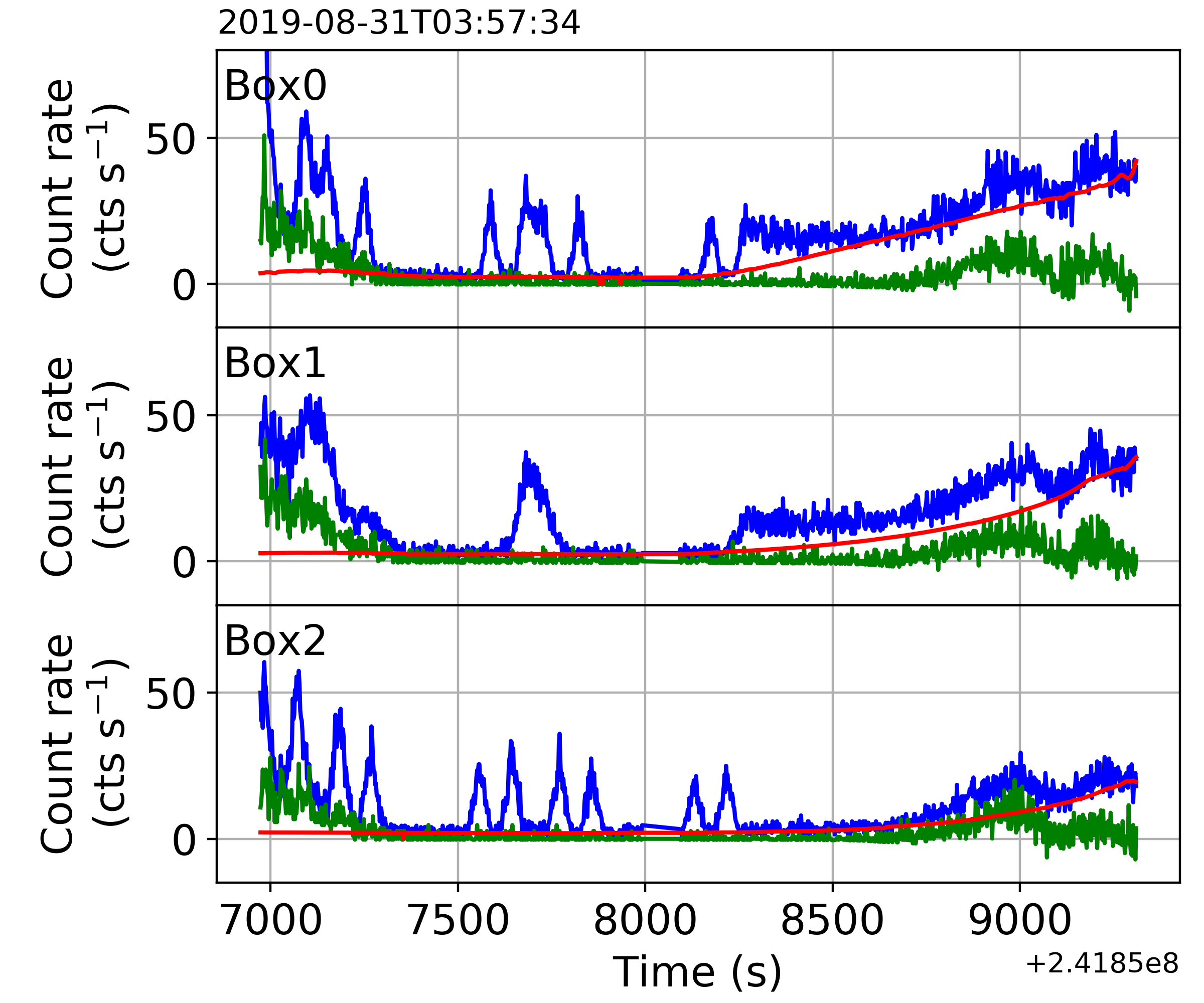}
\caption{LE light curves with the normal scanning and data download mode. The blue and green curves are light curves of $1-6$~keV and $8.6-10.3$~keV, respectively. The red curves are the backgrounds calculated by the SNIP method. \label{fig:lebkg}}
\end{figure}

In addition to the high background caused by particle events, there are also some special features in the scanning light curves. Figure \ref{fig:lebkg} shows the LE light curves with the normal scanning and data downloading mode and there is a count rate plateau in the right half of each panel. When the data downloading is performed, the observation continues but the scanning is suspended as the satellite stops rocking. Therefore, it is equivalent to a pointing observation at this time. As shown in Figure \ref{fig:lebkg}, the background during the data downloading is hard to be estimated accurately, thus the time interval for the satellite data downloading is also removed in the LE GTI selection. It is worth noting that the data downloading can also have a similar impact on ME and HE, thus the time intervals for the satellite data downloading are also removed in the ME and HE GTI selections.

\subsection{Special Filter of ME Data} \label{subsec:megti}
ME is very sensitive to space environment. During the scanning observations, some complicated background noises can appear in the light curves. The criteria of the ME GTI selection are stricter than these of LE and HE; however, it is still insufficient for some special cases shown as follows,

\begin{enumerate}
\item Particle event: similar to that in LE.
\item  Satellite data download: similar to that in LE.
\item In and out of SAA: the ME background is high and unstable near SAA, even lasting for $>600$~s after the passage from SAA. 
\item Unexpected sudden rise of the background in a single detector box.
\end{enumerate} 

Both the anomalous peaks of the cases 1 and 2 can be processed in the same way as LE (Section \ref{subsec:legti}); however, the anomalous peaks of the cases 3 and 4 often appear in the light curves with only one ME FOV, and it is difficult to distinguish whether these peaks are caused by particle events or scanned sources. The light curves without removing the unidentified peaks can be fitted directly, and the peaks may be fitted as new transient candidates. If the fitting result is consistent with these in LE and HE, the peaks caused by the scanned new transients and the candidates can also be confirmed, otherwise the peaks will be considered as anomalous peaks.

\begin{figure}[htbp]
\flushleft
\includegraphics[width=8.5 cm]{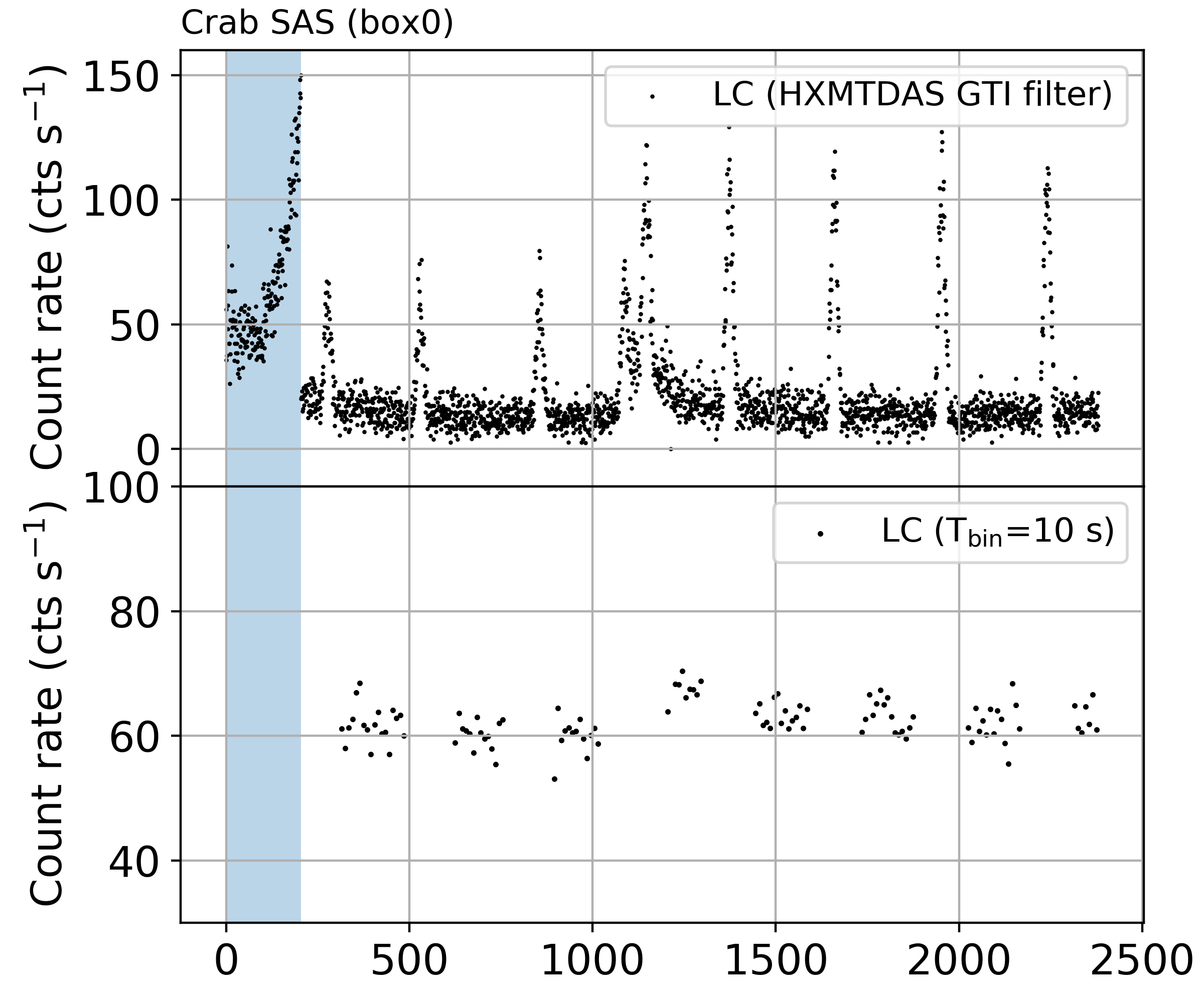}
\caption{The top panel shows the light curve with preliminary GTI filters; the bottom panel shows the light curve without all the peaks (source peaks $\&$ anomalous peaks). The shadow at the left end of the figure marks the time period of an anomalous peak. \label{fig:tct}}
\end{figure}

The most special feature of the ME data reduction is the pixel filter that will be described in the next subsection. Before the ME pixel filter, the clean background light curves without any peaks have to be obtained (see Section \ref{subsec:me pixel} for details). As shown in the top panel of Figure \ref{fig:tct}, there are a lot of peaks in a ME scanning light curve after the preliminary data reduction. Most of the peaks are caused by the scanned sources; however, there are still some anomalous peaks in the light curve. Unlike the particle events of LE, the anomalous peak only appears in the light curve with one ME FOV. All the peaks including the source peaks and anomalous peaks are removed to obtain the net background net light curves, which can be used in ME pixel filter (the bottom panel of Figure \ref{fig:tct}). 

\subsection{ME pixel filter} \label{subsec:me pixel}
\begin{figure}[htbp]
\flushleft
\includegraphics[width=8.5 cm]{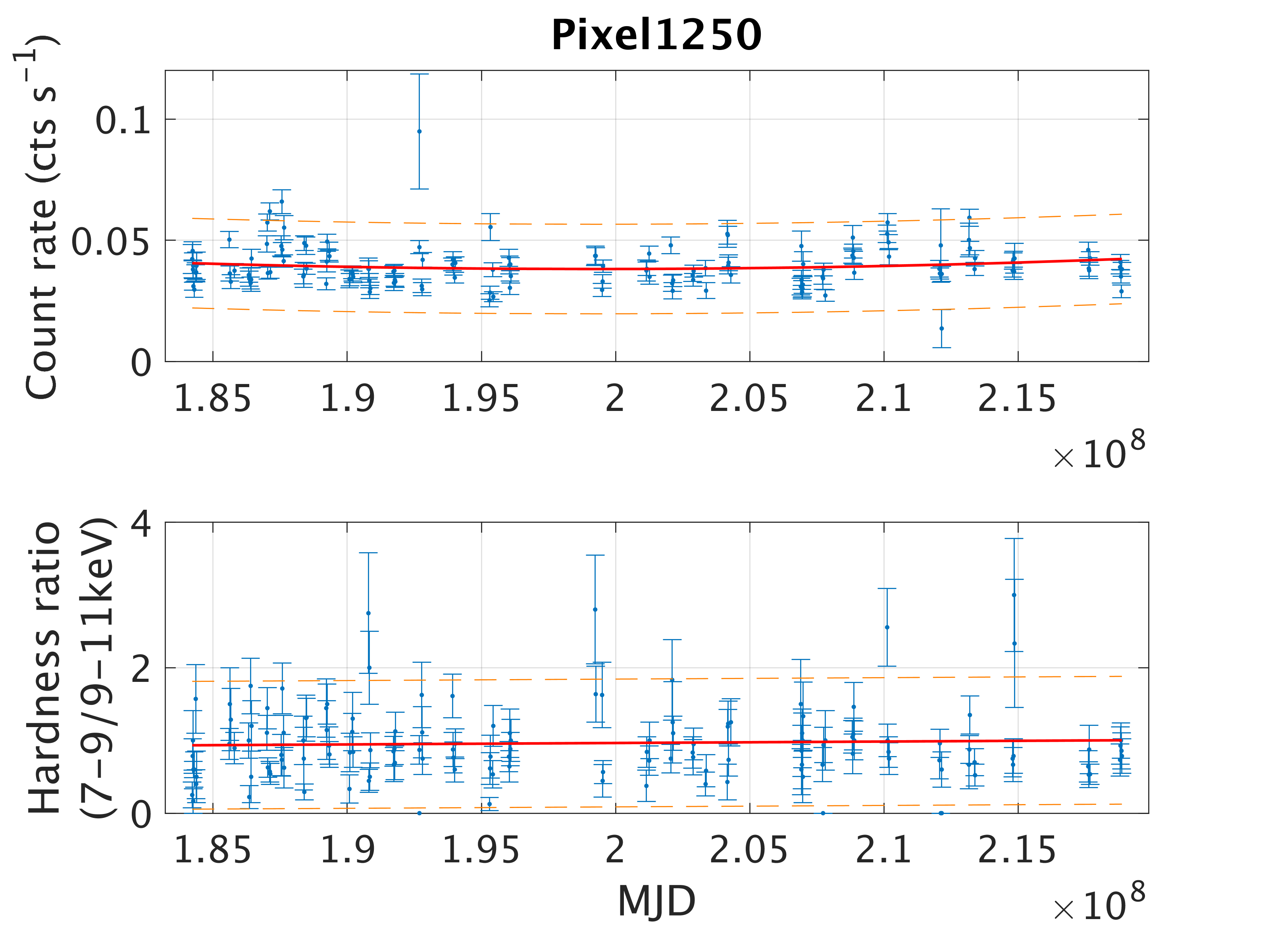}
\caption{Long-term light curve and hardness curve of a typical pixel (blank sky). This figure also shows the fitting functions of count rate and hardness ratio (blue point: blank sky data, red solid line: ideal count rate curve, orange dashed line: ideal count rate limit)\label{fig:pixelf}}
\end{figure}

With the 141 pointing observations of the blank sky regions before 
January, 2019, the long-term light curves of every ME pixel in $5-40$~keV are produced and then fitted with the empirical functions. Most of the fitting functions are the second-order polynomials, and some are third-order polynomials and piecewise functions. The fitting of the long-term hardness ratio ($\rm 7-9~keV/9-11~keV$) are also performed and the fitting results are obtained as an auxiliary reference. Figure \ref{fig:pixelf} shows an example of the long-term light curve and hardness ratio fitting of a ME pixel. The data is fitted with a second-order polynomial and the standard deviation $\sigma$ of the fitting residuals is obtained. According to the fitting results of all pixels and the known bad pixels catalog, all the ME pixels are classified into seven groups. The detailed classification criteria and the results are shown in Table \ref{tab:pixlevel}. 

As shown in Table \ref{tab:pixlevel}, the number of available pixels (top four groups) is more than 1300 on average. We thus use the data from the pixels with small FOVs in the top four group of Table \ref{tab:pixlevel}. As described above, the ME pixels vary both in the detection efficiency and background. Therefore, the pixel filter must be performed in the data reduction of each scanning observation. The average count rate of the background $R$ is obtained from the light curve without the cataloged source signal. As shown in Figure \ref{fig:pixelf}, the dashed lines above and below the model line are the $3\sigma$ range that is considered as the background's effective range of the pixel in the normal state. If $R$ is within the background's effective range, the pixel is considered as a regular pixel in this scanning observation.

The ME pixel filter method works well in the scanning observations of \emph{Insight-HXMT}. Generally, the number of bad pixels is about $420-480$, and thus more than 1000 pixels with the small FOVs can be used in the scanning observations.

\begin{deluxetable*}{ccl}[htbp]
\tablecaption{ME pixel classification by blank sky scanning \label{tab:pixlevel}}
\tablecolumns{8}
\tablenum{3}
\tablewidth{0pt}
\tablehead{
\colhead{Level} &
\colhead{Number} &
\colhead{Explanation}
}
\startdata
1 & 1121 & Good pixels: count rate dispersion $\leqslant 0.05$. \\
2 & 107 & Pixels that can be used: count rate dispersion between 0.05 and 0.06. \\
3 & 15 & Pixels that can be used: count rate has a step, and the segmentation fit of each part $\leqslant 0.06$. \\
4 & 117 & Pixels that can be used only in some cases: count rate dispersion between 0.06 and 0.10. \\
5 & 36 & Not recommended pixels: count rate dispersion $\geqslant 0.10$, and 5 pixels fitted with a cubic function. \\
6 & 119 & Unavailable pixels: pixels with abnormal count rate, radio sources, and some problem pixels. \\
7 & 214 & Closed pixels. \\
\enddata
\end{deluxetable*}

\section{Light curve analysis process} \label{sec:lc analyse}
\subsection{PSF model} \label{subsec:psf}
\begin{figure}[htbp]
\centering
\includegraphics[width=5 cm]{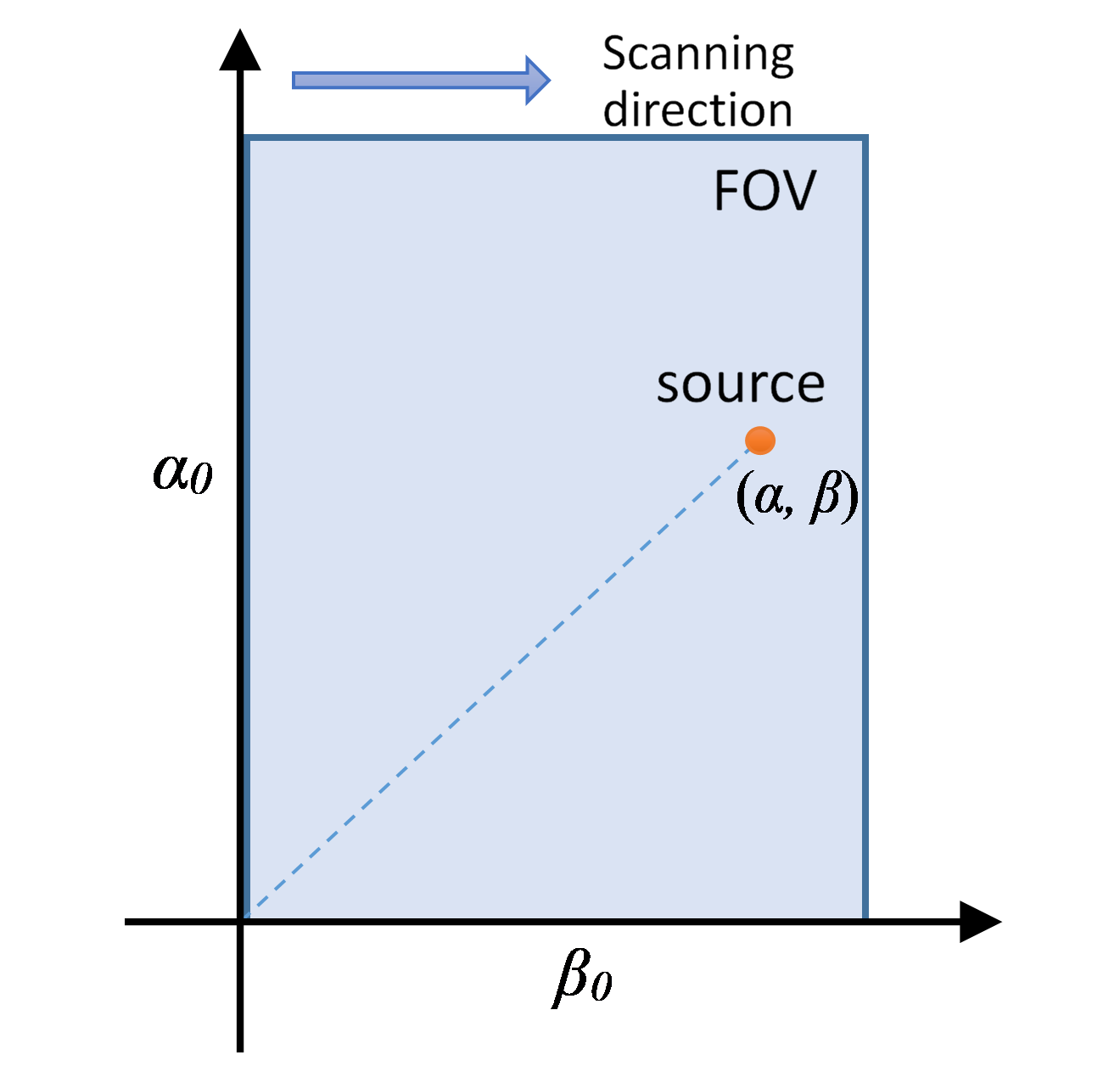}
\caption{Schematic diagram of the orientation angles of a source in FOV. \label{fig:psf}}
\end{figure}

With the progressive scanning mode described in Section \ref{sec:intro}, the sources in the scanned area cross the FOV several times. From the light curves of the scanning observations of \emph{Insight-HXMT}, we can see various triangular peaks that are the signals as the sources cross the FOV. Moreover, the efficiency is the function of the orientation angles of the FOV, which can be analogized as the point-spread function (PSF) of the collimating telescope. For the collimator of \emph{Insight-HXMT}, the geometrical PSF can be expressed as
\begin{equation}
f(\alpha,\beta)=C\frac{(1-\frac{abs(\tan(\alpha))}{\tan(\alpha_0)})(1-\frac{abs(\tan(\beta))}{\tan(\beta_0)})}{\sqrt{\tan^2(\alpha)+\tan^2(\beta)+1}},
\label{equ}
\end{equation}
where $\alpha$ and $\beta$ are the orientation angles of sources relative to the major axis and minor axis of the telescopes respectively, $\alpha_0$ and $\beta_0$ are the sizes of long side and short side of FOV (FWHM) (Figure \ref{fig:psf}), \emph{C} is the normalized flux as the source at the center of the FOV. All the collimators of the three telescopes have rectangular FOVs (Figure \ref{fig:fov}) with different sizes of FOV and orientations, and can be described by the geometric PSF model as equation \ref{equ}. The in-orbit calibration of the geometrical PSF with rotation correction and parabolic correction are performed by \citet{2019N}, and the light curve fitting is done with the calibrated PSF in order to obtain more accurate analysis results.

\subsection{Source catalog} \label{subsec:catalog}
The source catalog is a combination of \emph{Swift}, \emph{INTEGRAL} and MAXI catalogs from their websites. If some sources are close to each other, their peaks on the light curve will overlap. This can result in the highly uncertain source fluxes in light curve fitting, which are difficult to be distinguished by the scanning observations of \emph{Insight-HXMT}. For the scanning observations to the Galactic center, there are many sources in a scanned area. The sources within $0^{\circ}.3$ are merged as one source in the joint-fitting of the light curve of LE. Due to the low source flux of ME and the high background intensity of HE, the degeneracy of the adjacent sources are more serious in ME and HE than that in LE, and the merge ranges are $0^{\circ}.6$ for both ME and HE. Finally, the source number in the catalogs of LE, ME, and HE are 2879, 2288 and 2288, respectively.

\subsection{Light curve fitting} \label{subsec:lc fitting}
For each telescope, the light curves with the three FOVs are obtained and then fitted jointly. With the source catalog of each telescope, the degeneracy of the adjacent sources is avoided. The analysis process of the scanning light curve can be divided into the following five steps:

\begin{figure}[htbp]
\centering
\plotone{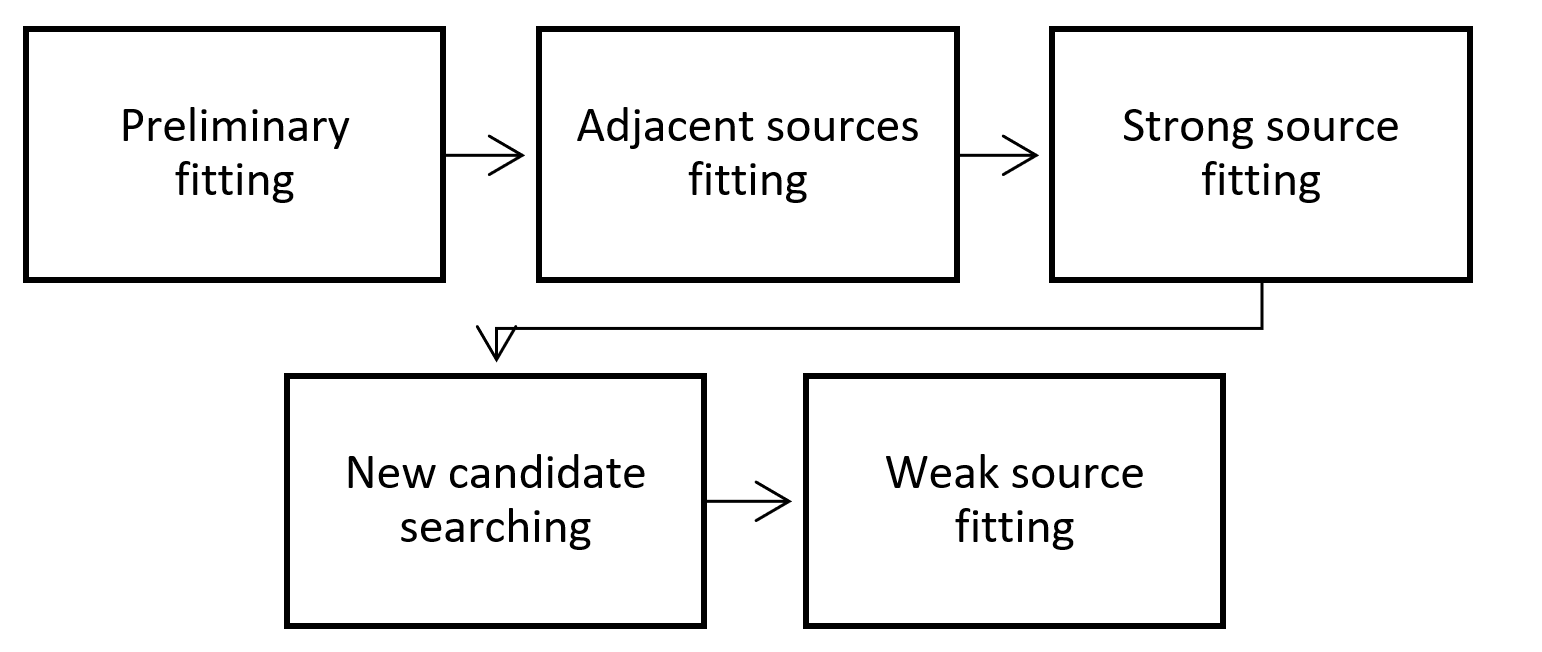}
\caption{Analysis process of the scanning light curve. \label{fig:lcprocess}}
\end{figure}

\begin{enumerate}
\item Make the classification to the known sources according to Signal to Noise Ratio (SNR).

Each source in the scanned area is fitted separately to obtain an approximate SNR. All the sources with $\rm SNR > 2$ are classified as strong sources and that with $\rm SNR < 2$ are classified as weak sources.
\item Fit the light curve only with the strong sources. 

All the strong sources selected in step 1 are fitted simultaneous with fixed positions and free fluxes. 
\item Fit the residuals of step 2 with a hypothetical source. 

A hypothetical source is used to check whether there is an unknown source not in the source catalog. In this fitting, both the position and flux are free parameters. We search for the unknown source throughout the entire scanning area. In order to avoid dropping into a local optimal solution, we perform multiple fittings with different initial positions along the scanning trace. The parameters with the smallest $\chi^2$ are regarded as the final result.
\item Make the judgment of the hypothetical source.

If the SNR is more than 5, the hypothetical source will be considered as a new transient candidate that need to be identified further to reduce the probability of false alarms. 
\item Fit the residuals of step 3 with the weak source selected in step 1. 

Each weak source is fitted separately with the position fixed to that in the catalog and the flux as a free parameter. 
\end{enumerate}

\begin{figure*}[htbp]
\gridline{
	\fig{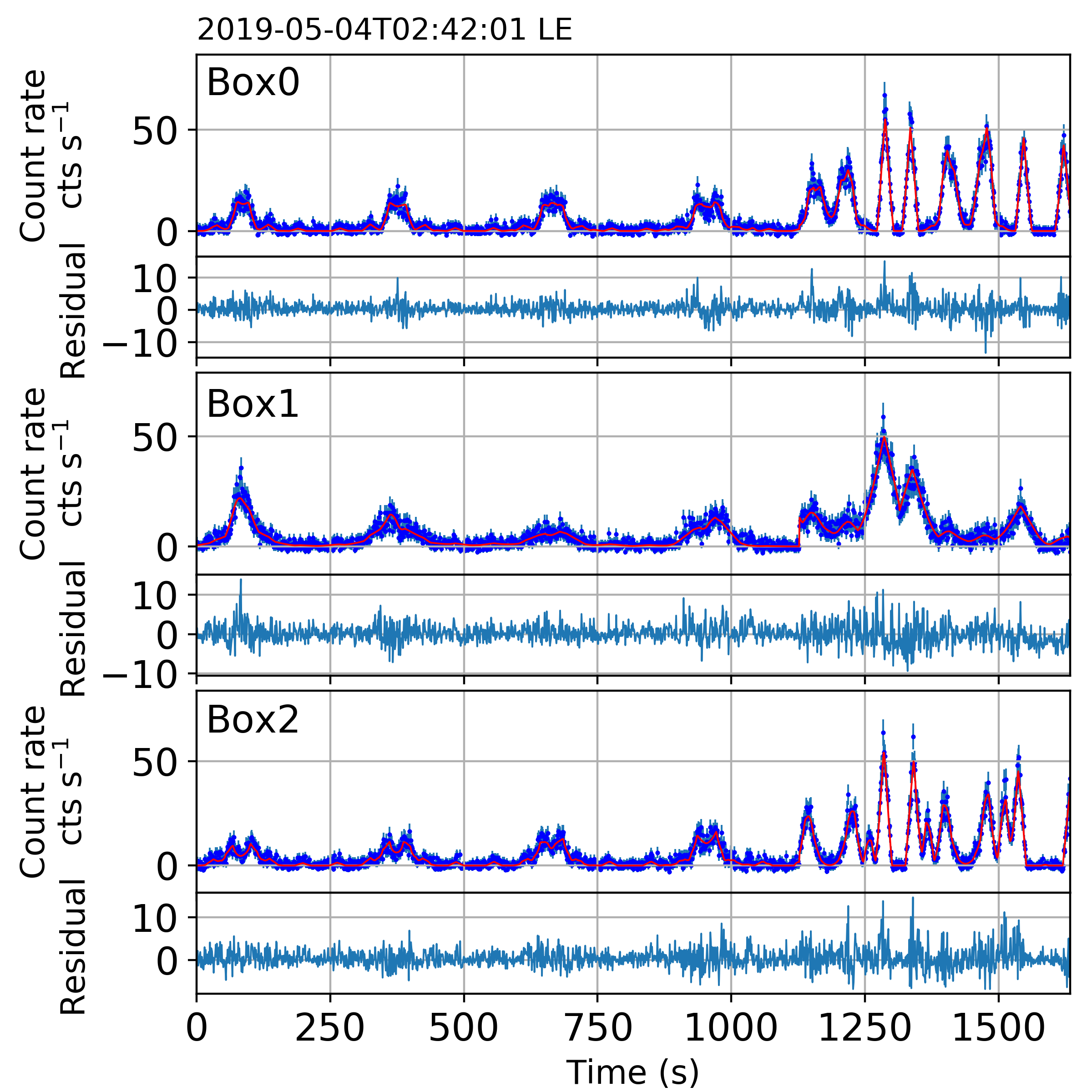}{0.45\textwidth}{}
	\fig{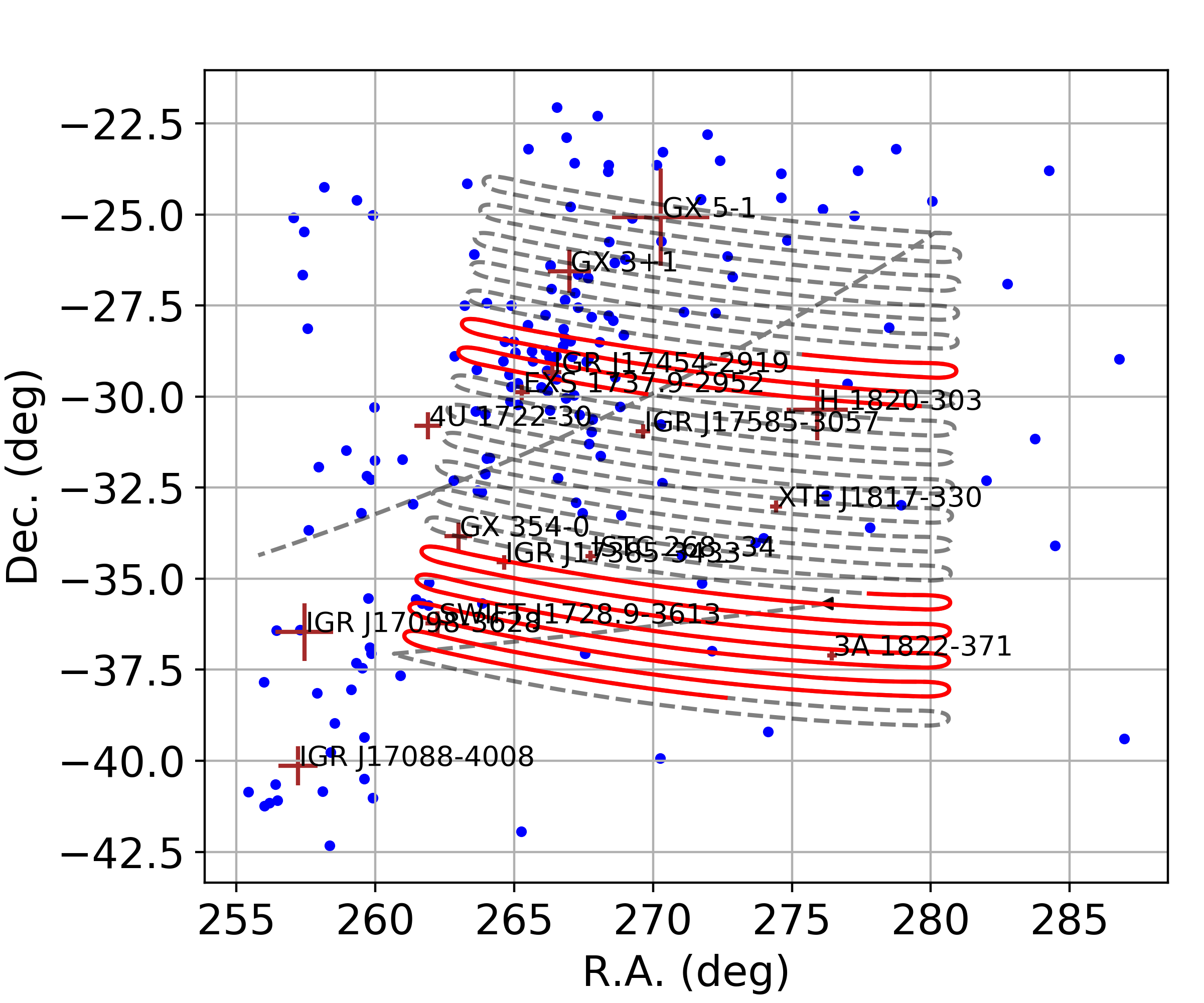}{0.45\textwidth}{}}
\caption{An example of light curve fitting and scanned area of LE. In the left panel, the red lines are the model curves. In the right panel, the blue points are weak sources and the red crosses are strong sources; the red solid lines are scanning tracks with GTI and the grey dashed lines are the scanning tracks in the whole scanning observation.\label{fig:fitle}}
\end{figure*}

\begin{figure*}[htbp]
\gridline{
	\fig{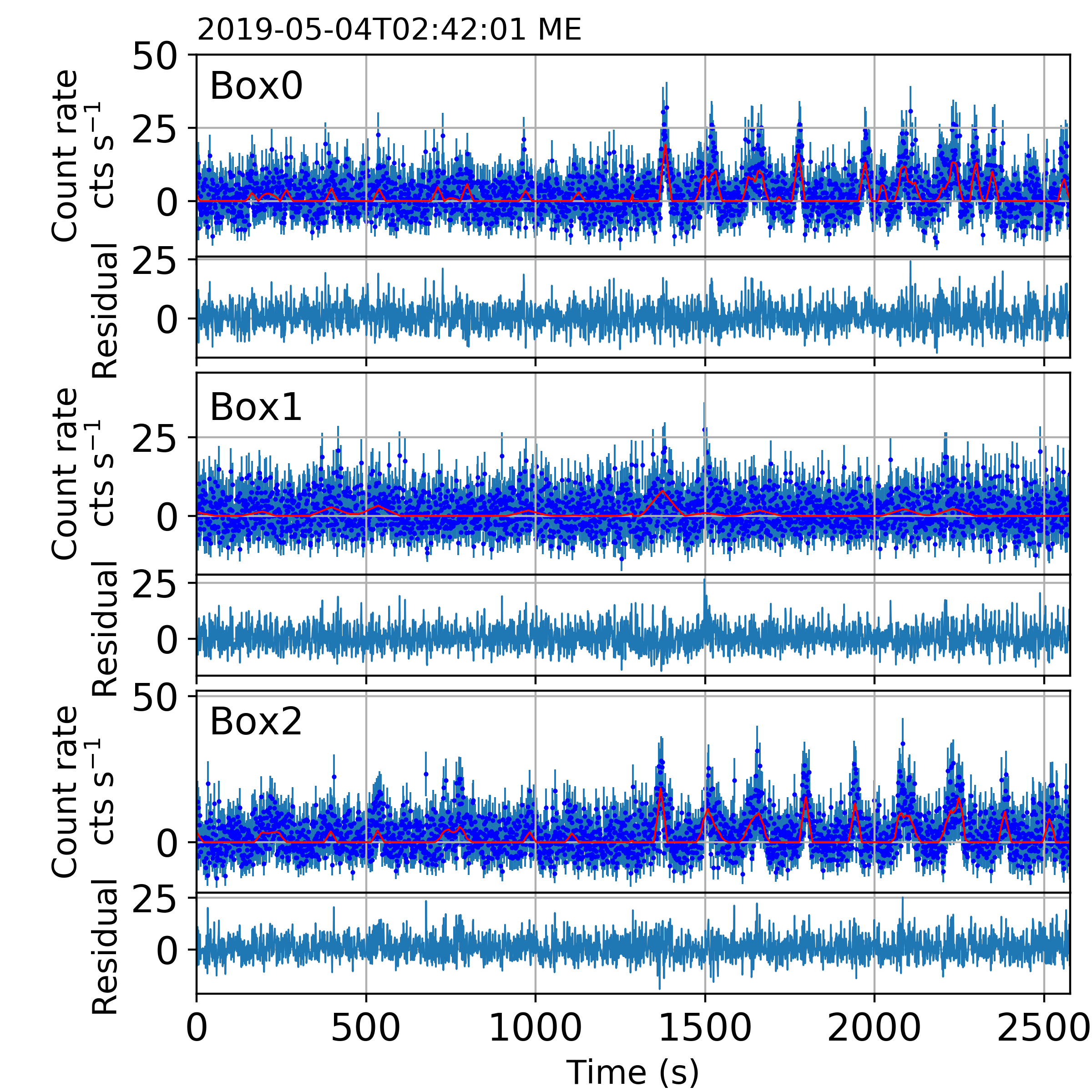}{0.45\textwidth}{}
	\fig{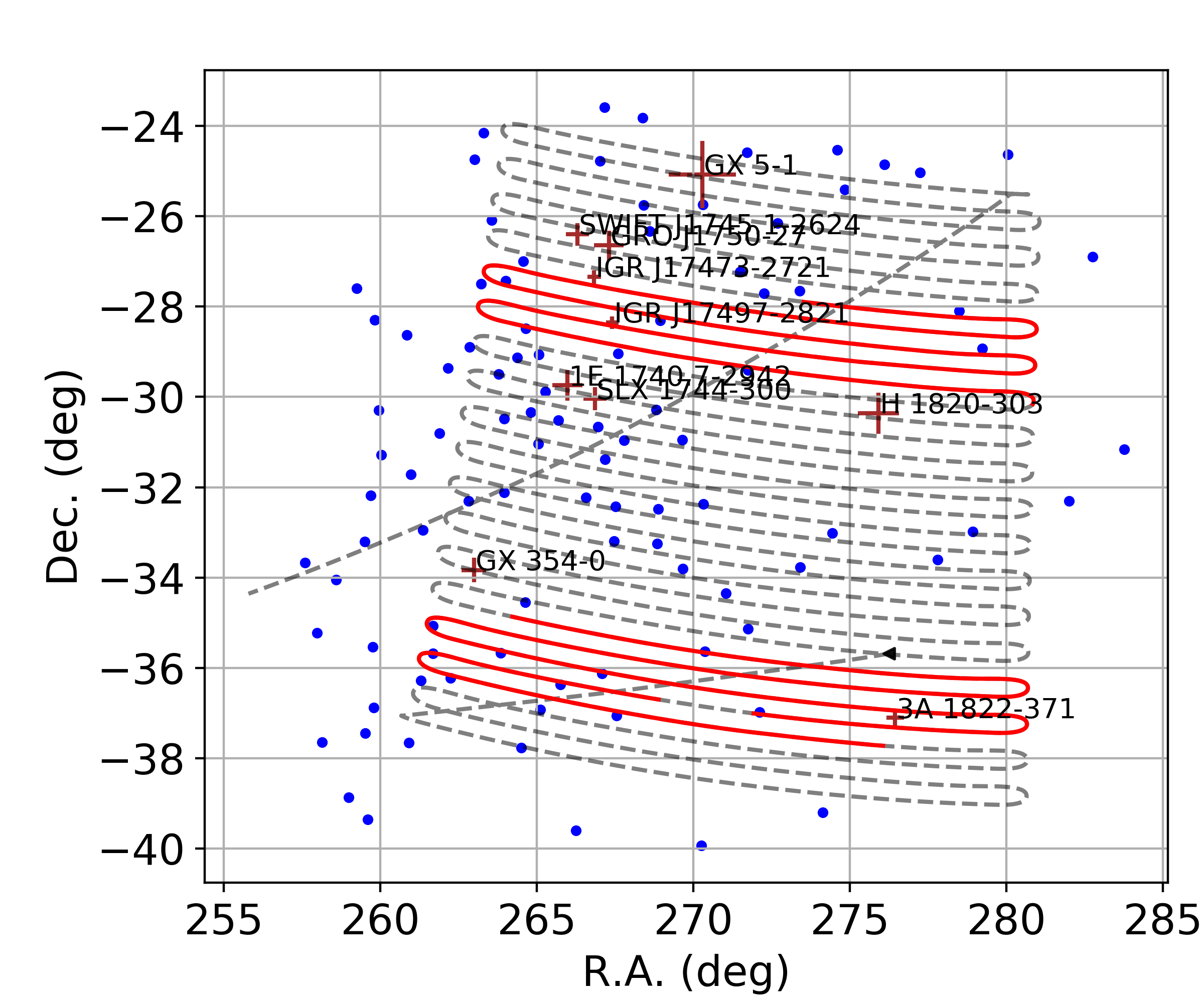}{0.45\textwidth}{}}
\caption{The same as Figure \ref{fig:fitle}, but for ME. \label{fig:fitme}}
\end{figure*}

\begin{figure*}[htbp]
\gridline{
	\fig{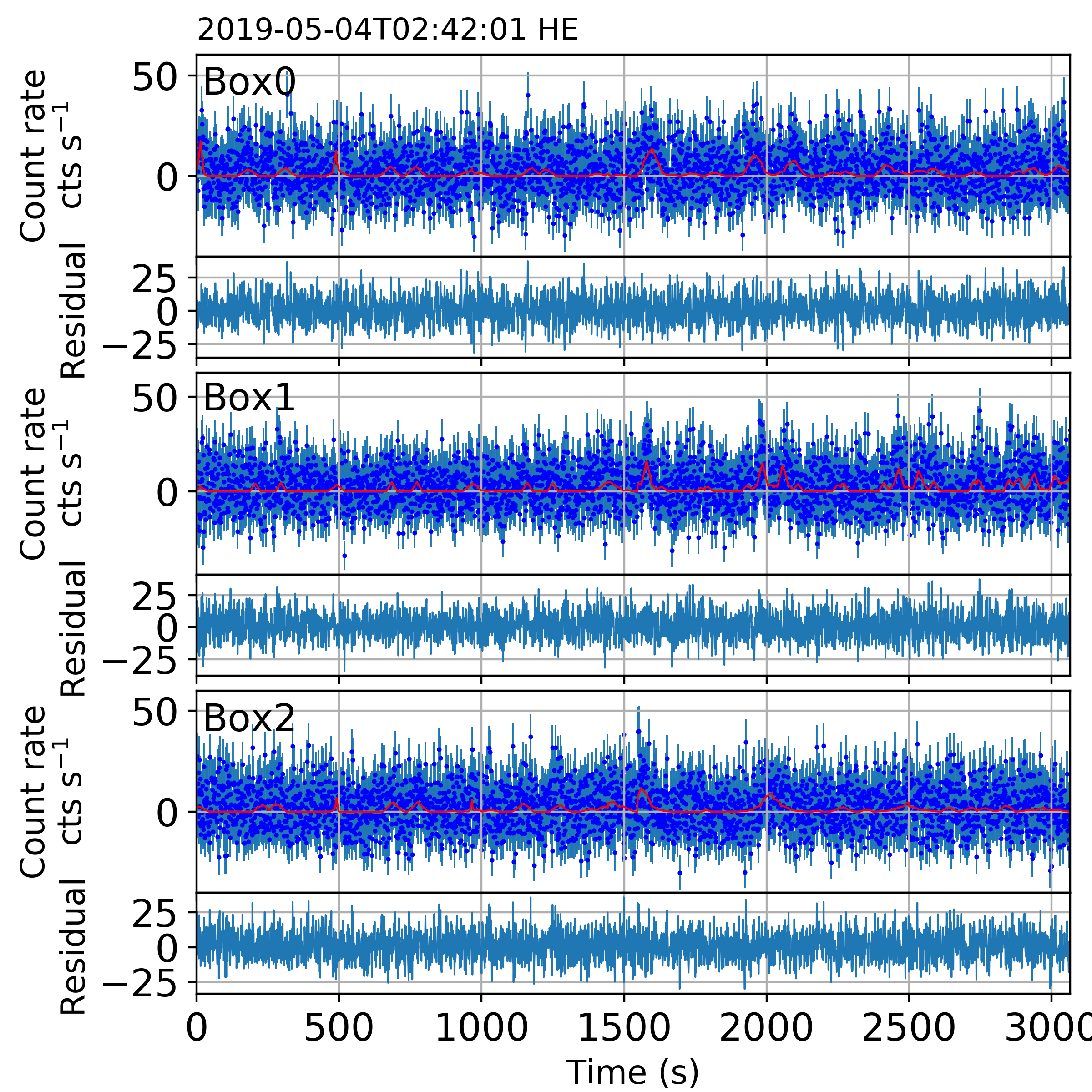}{0.45\textwidth}{}
	\fig{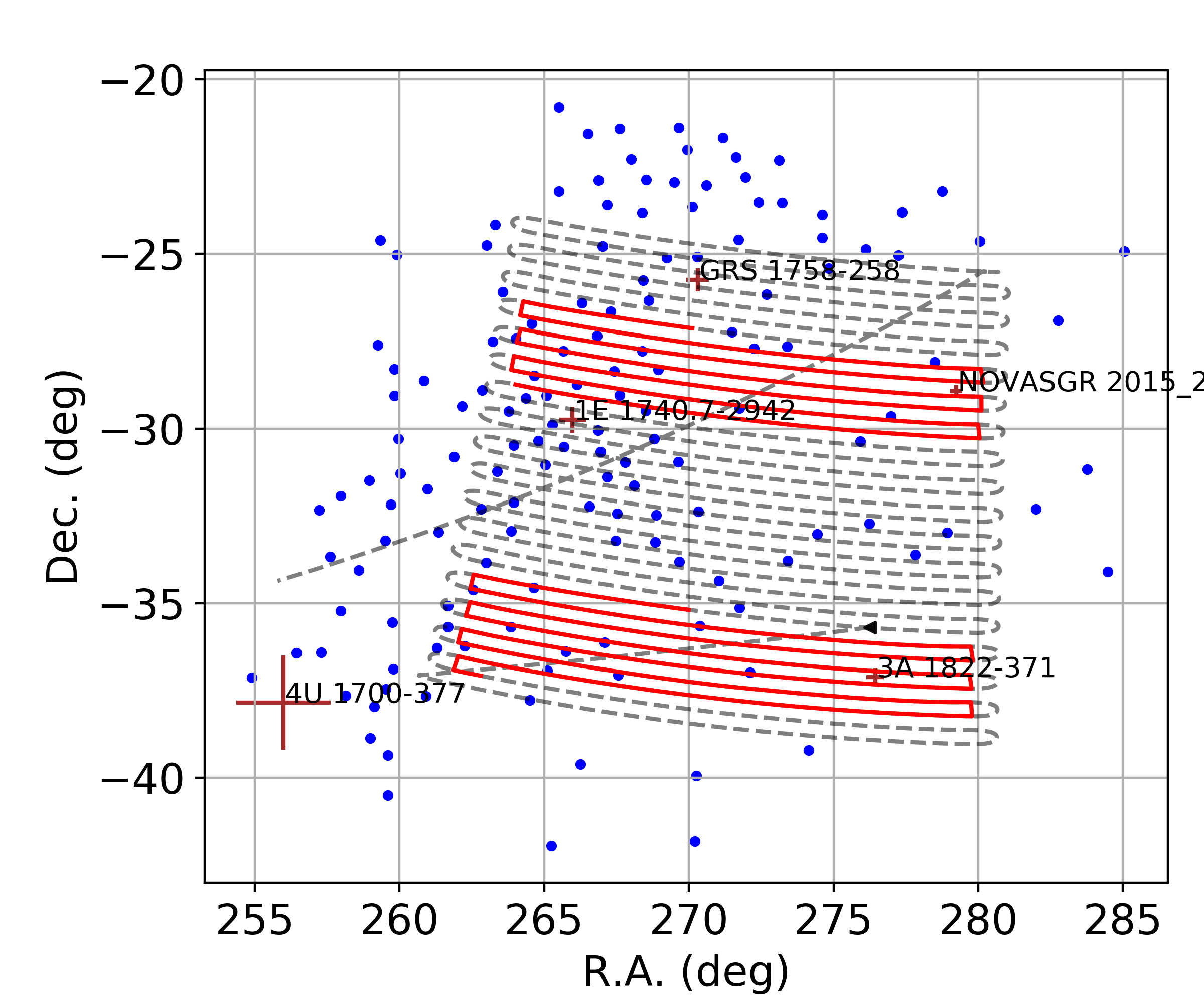}{0.45\textwidth}{}}
\caption{The same as Figure \ref{fig:fitle}, but for HE. \label{fig:fithe}}
\end{figure*}

For a scanning observation of \emph{Insight-HXMT}, the fluxes of the strong sources can be obtained by the above steps. If a new X-ray transient appears in the scanned area, the position and flux can also be obtained. Figure \ref{fig:fitle}$-$\ref{fig:fithe} are examples of the scanning light curve fitting of all three telescopes. If the signal of a new transient candidate with low significance (i.e., SNR~$<$~5) appears in the same sky area multiple times, we will consider it as a candidate and will take further identification such as \emph{Insight-HXMT} deep scanning observation and possibly also ToO observation with \emph{Swift}/XRT for imaging verification.

\section{Result} \label{sec:result}
\begin{figure*}[htbp]
\centering
\includegraphics[width=12 cm]{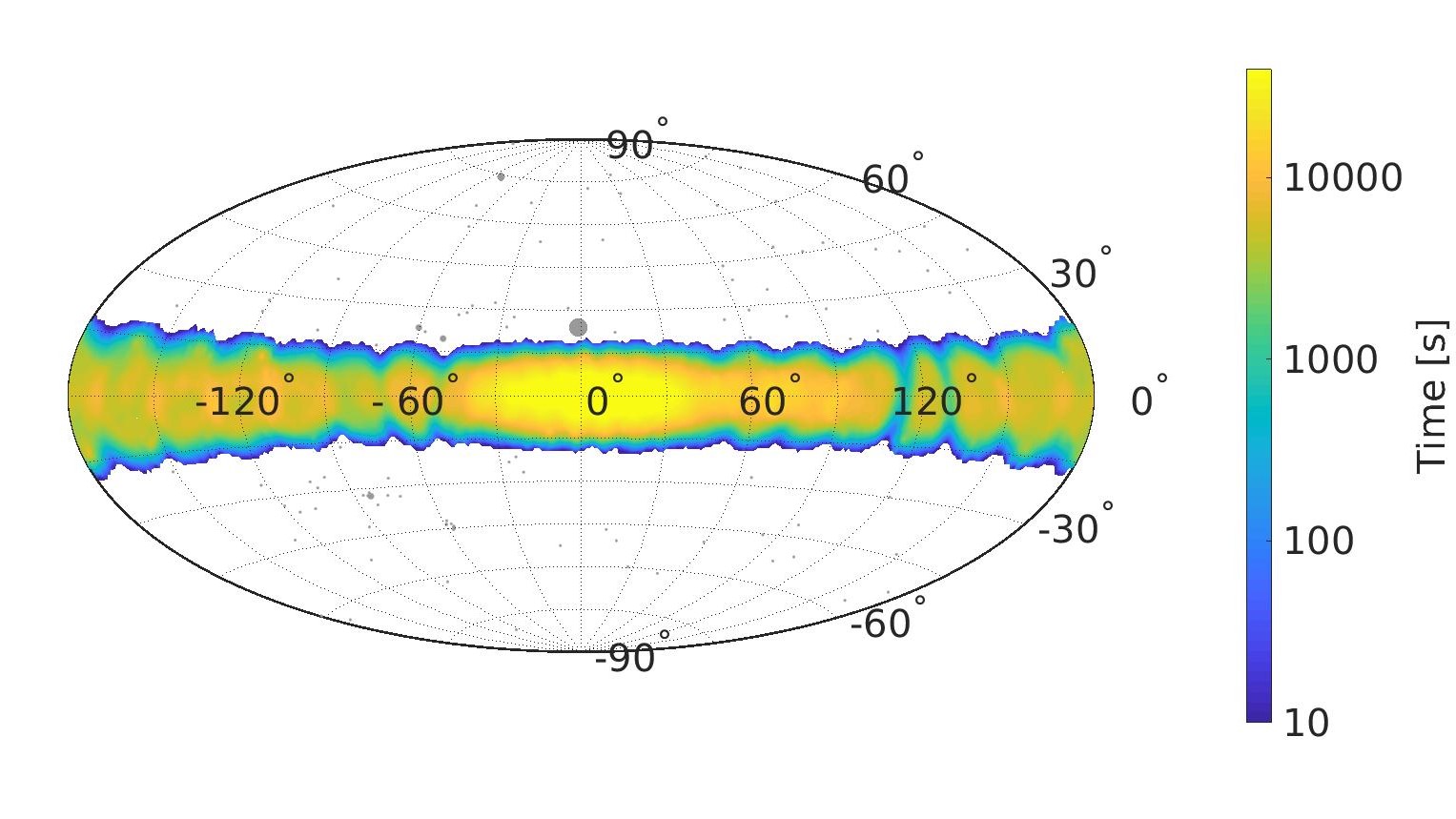}
\caption{Exposure sky map for the Galactic plane scanning
survey up to 2019-10-24.\label{fig:expo}}
\end{figure*}

\emph{Insight-HXMT} has performed more than 1000 scanning observations covering the whole Galactic plane up to September 2019. From the exposure sky map shown in Figure \ref{fig:expo}, it can be seen the scanning observations are performed more on the Galactic center in the Galactic plane scanning survey. 

By now, the whole Galactic plane has been covered completely. As described in Section \ref{sec:intro}, there are several scanning parameter groups that can be used in scanning observations. Before the end of April, 2019, the whole Galactic plane is divided into 22 areas with radius $\rm \emph{R}=10^{\circ}$. As the scanning mode shown in Figure \ref{fig:scan}, the scanning observation is performed with $\rm \emph{v}=0^{\circ}.06~s^{-1}$ and $\rm \emph{d}=0^{\circ}.1$. We find that the GTI of LE is usually less than $\rm 3~ks$ and sometimes fragmented. In order to increase the probability of discovering new and weak sources appearing in the scanned region, the whole Galactic plane is redivided into 50 areas with $\rm \emph{R}=7^{\circ}$ and the scanning interval is adjusted to $\rm \emph{d}=0^{\circ}.4$ after the end of April, 2019. With the new scanning parameters, the LE telescope can scan one regions as many times as possible within a continuous GTI.

\begin{figure*}[htbp]
\gridline{
	\fig{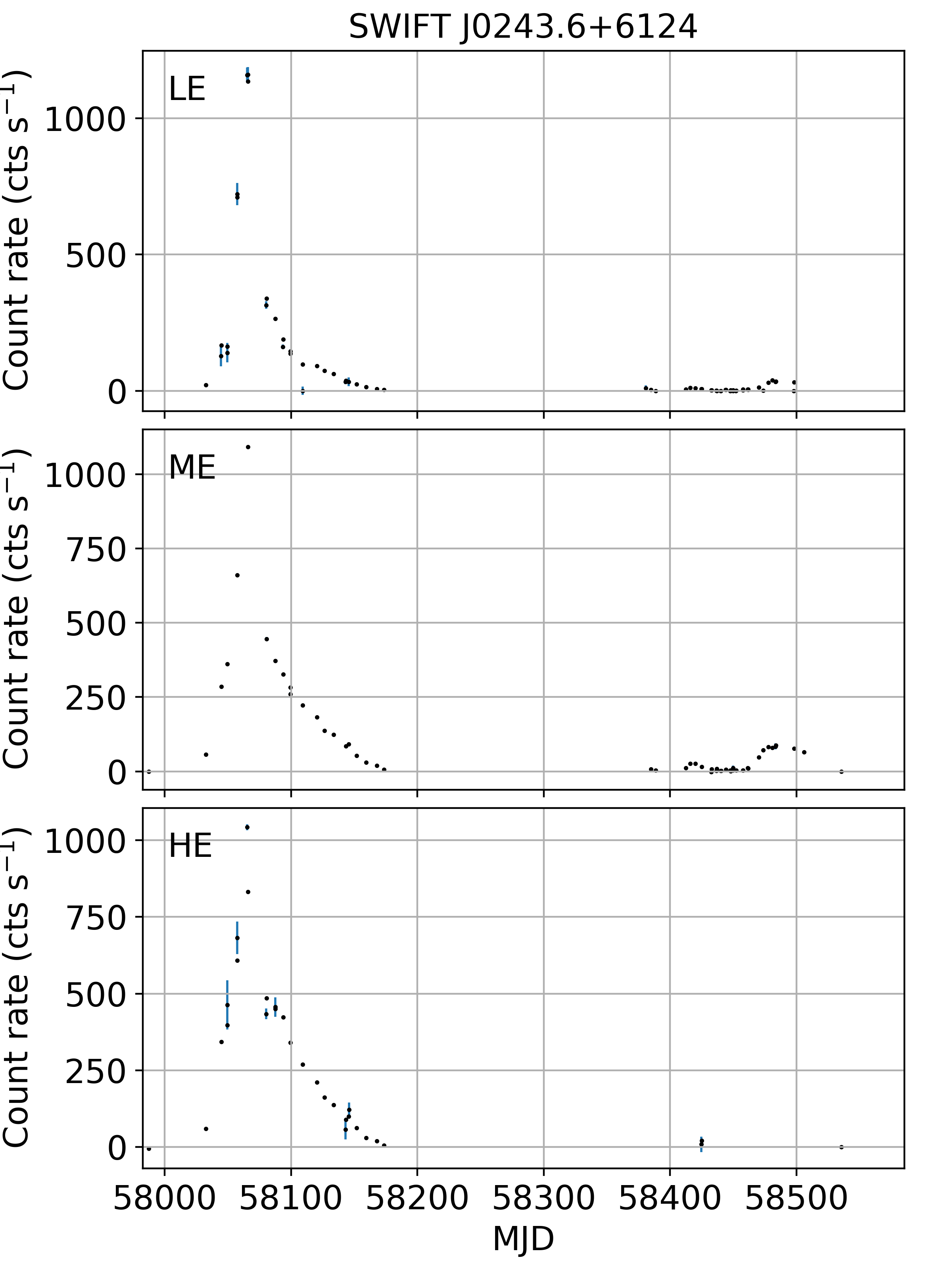}{0.4\textwidth}{(a)}
	\fig{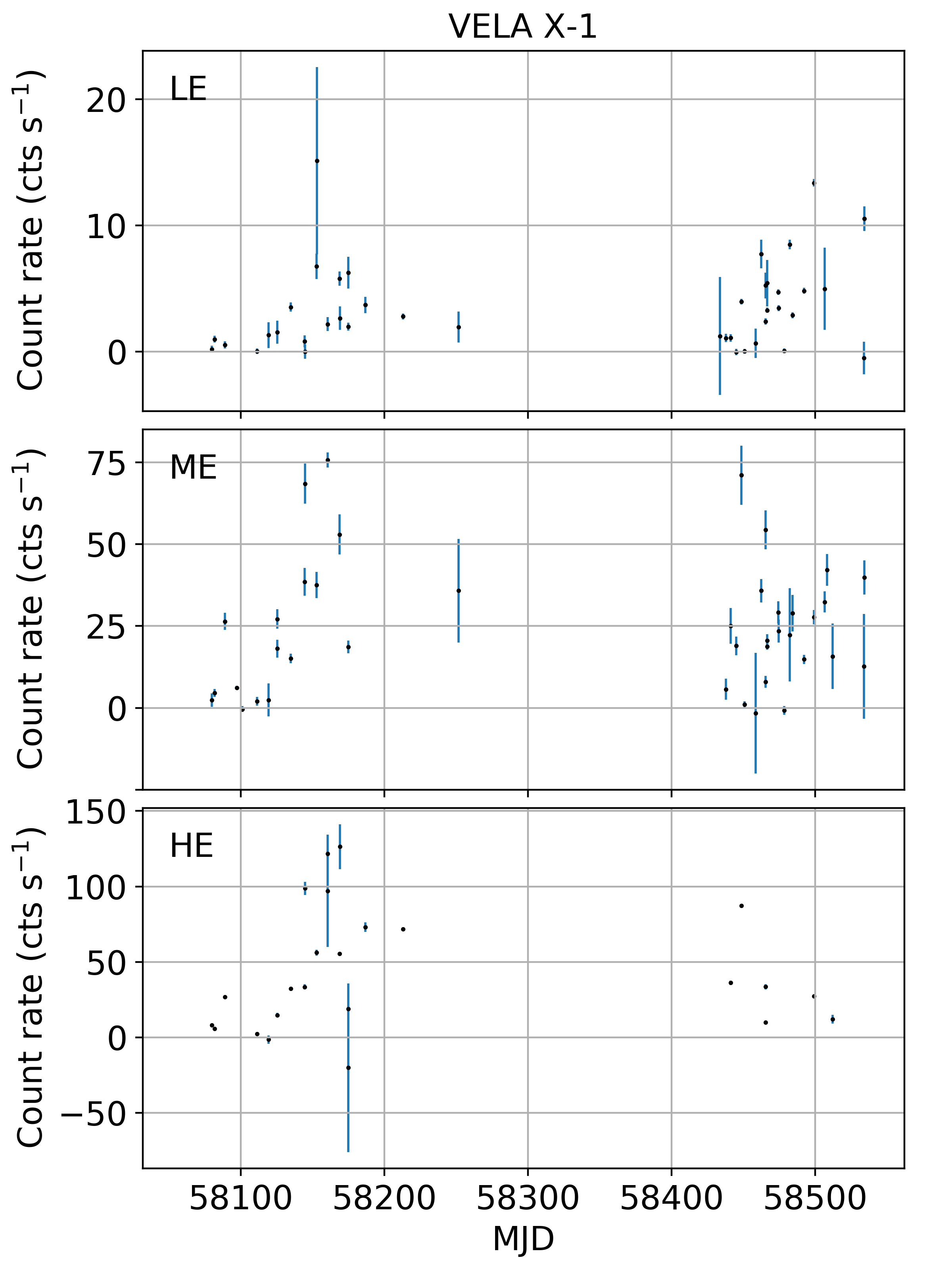}{0.4\textwidth}{(b)}}
\gridline{
	\fig{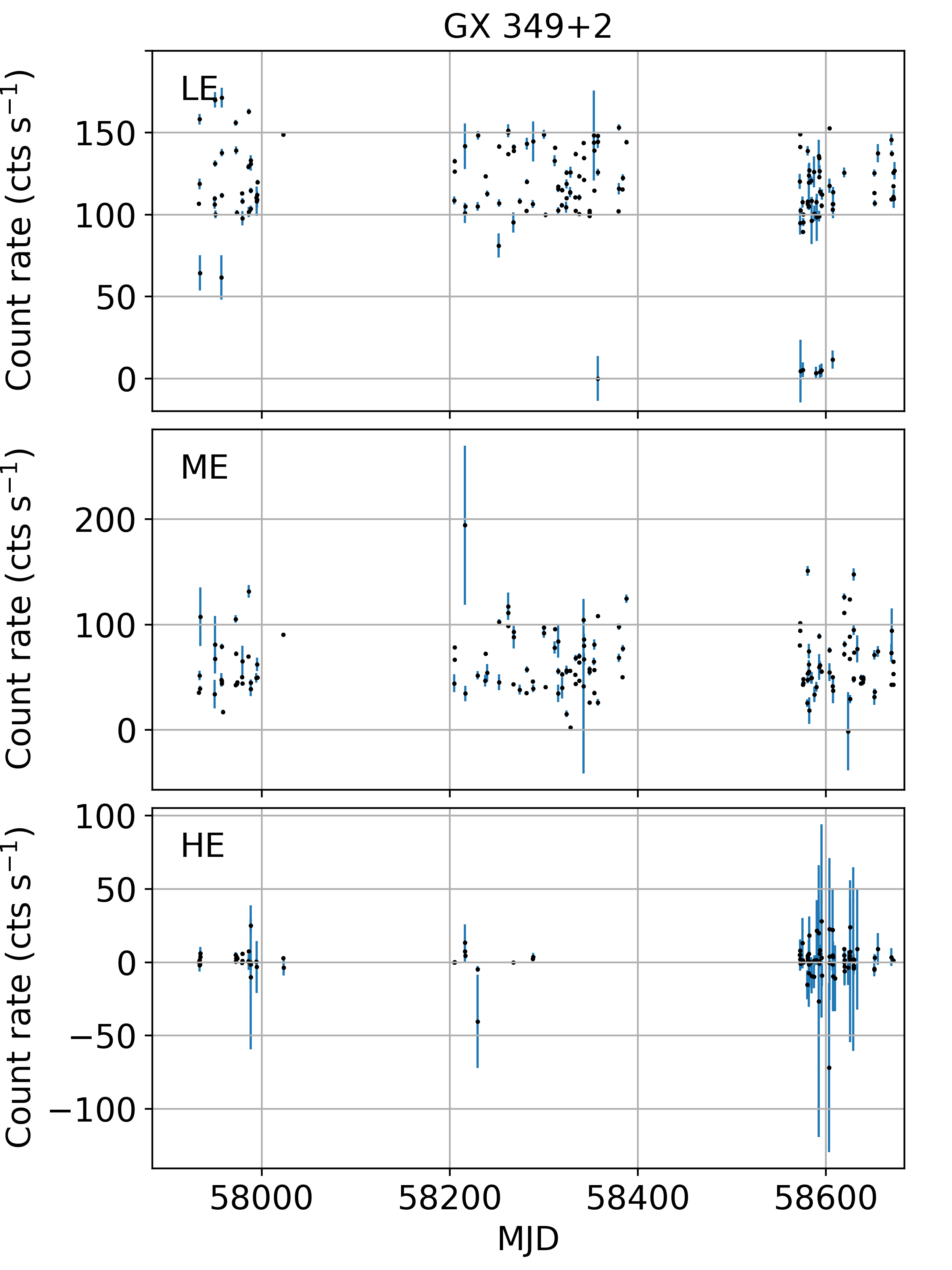}{0.4\textwidth}{(c)}
	\fig{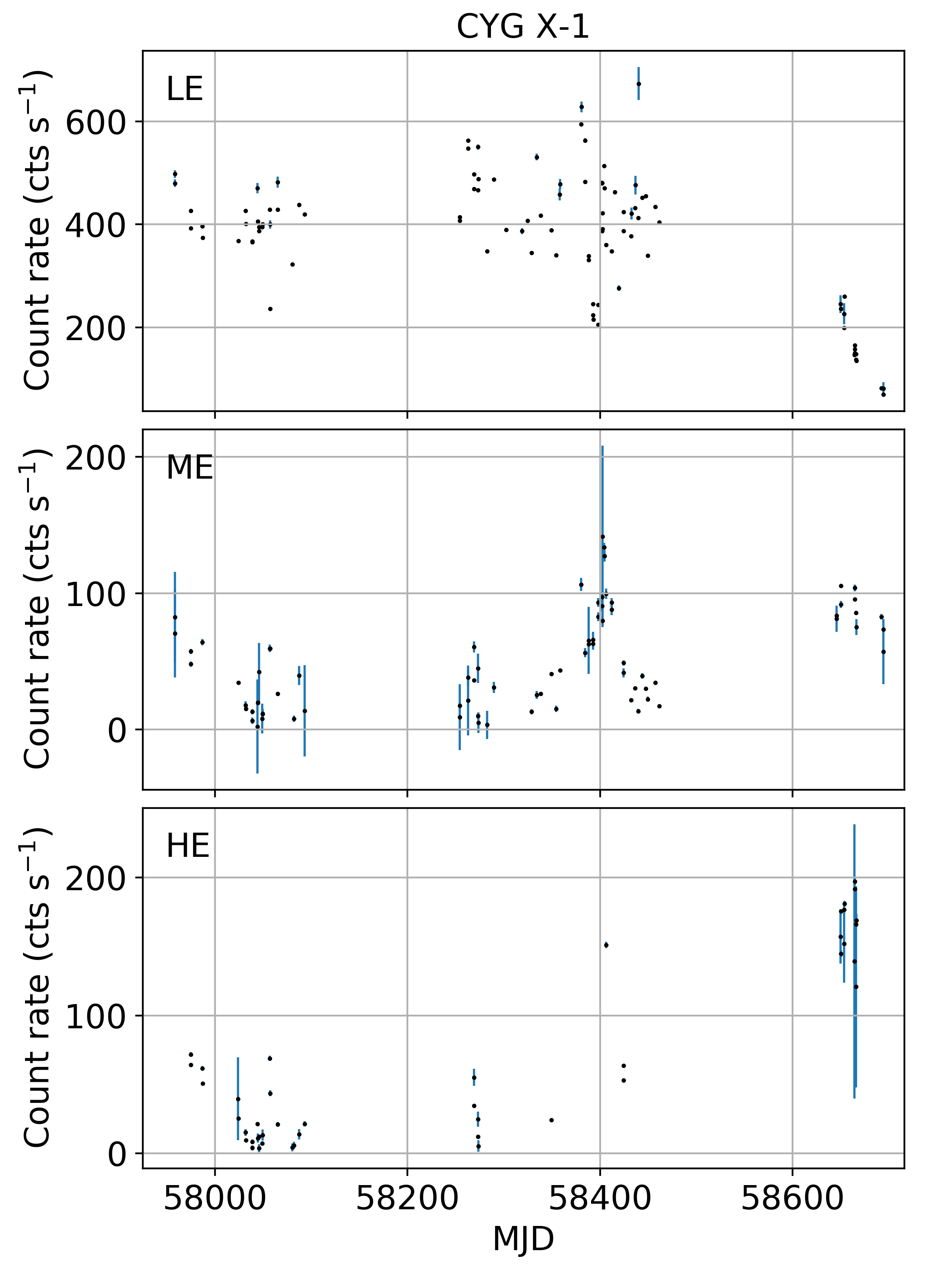}{0.4\textwidth}{(d)}}
\caption{Long-term light curves of Swift J0243.6+6124 (BeXRB), Vela X-1 (HMXB/NS), GX 349+2 (LMXB/NS) and Cyg X-1 (HMXB/BH). \label{fig:longtermlc}}
\end{figure*}

More than 800 known X-ray sources with various types are monitored by LE, ME and HE, respectively, and the long-term light curves with different energy ranges are obtained to do further scientific analysis. Since the spectral characteristic varies over time, the long-term light curves of the sources of the three telescopes are different with each other. Swift J0243.6+6124 is an X-ray binary consisting of a neutron star and a Be star, and it had a strong outburst during 2017 to 2018 \citep{2017ATel...10822, 2019ApJ...879...61}. \emph{Insight-HXMT} monitored the entire outburst completely and the long-term light curve in three energy ranges are shown in Figure \ref{fig:longtermlc} (a). In addition to Swift 0243+6124, the long-term light curve of Vela X-1 (HMXB/NS), GX 349+2 (LMXB/NS), Cyg X-1 (HMXB/BH) are also shown in Figure \ref{fig:longtermlc} (b)-(d) as examples.

\begin{figure*}[htbp]
\centering
\includegraphics[width=9 cm]{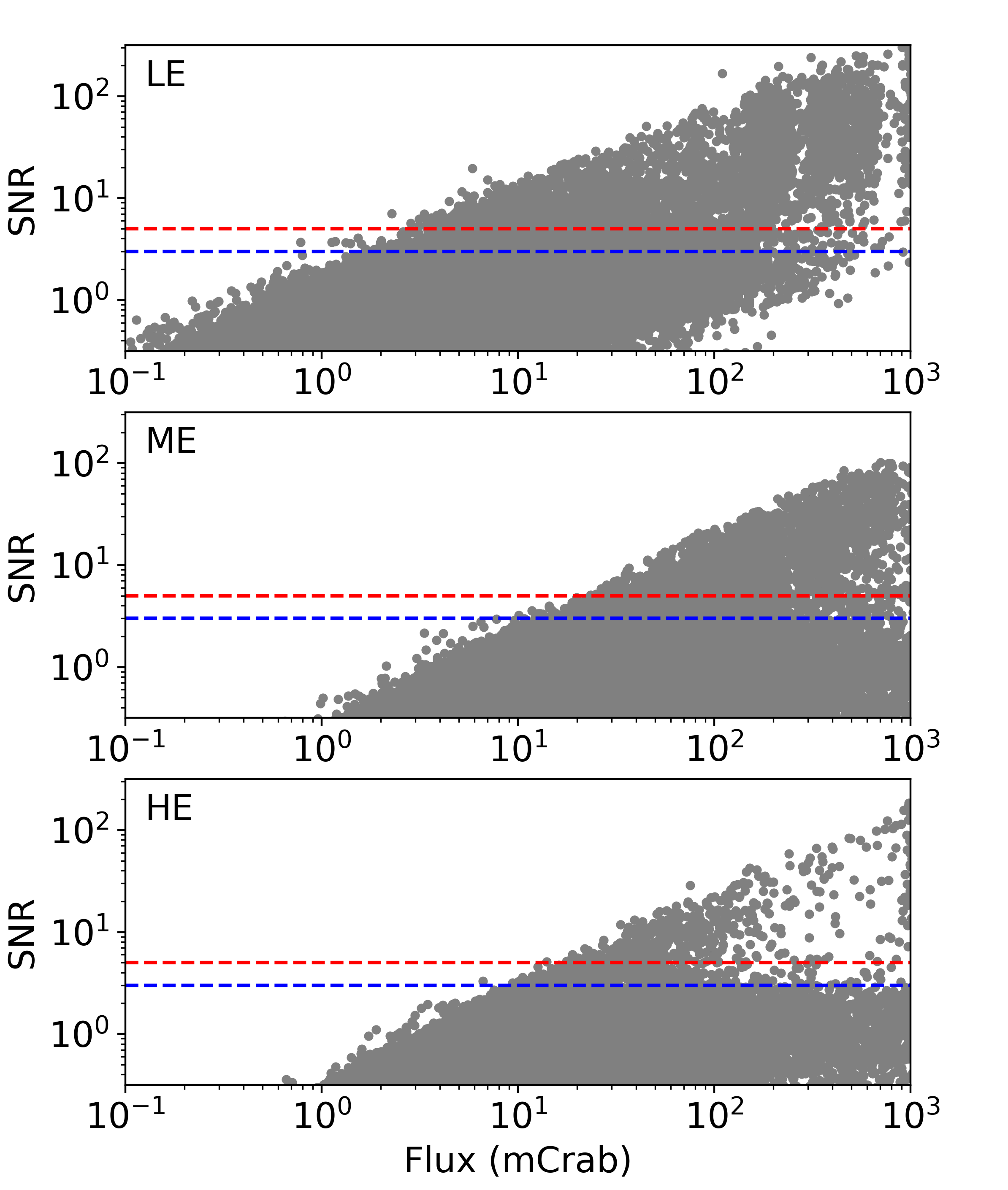}
\caption{Relationship of the flux and SNR of LE, ME and HE, respectively. The red and blue dashed lines are ${\rm SNR}=3$~and~5, respectively. \label{fig:sensi}}
\end{figure*}

From \citet{2019N}, the systematic errors in flux estimation are $1.8\%$, $1.6\%$ and $2.7\%$ for LE, ME and HE (Table \ref{tab:sensi}), which may be caused by various factors, such as the small uncertainty of the Energy-Channel relation and the weak long-term evolution of the PSF. As the spectral characteristic of the Crab is generally stable, the long-term light curves of the Crab can be used to obtain the systematic errors of the estimated flux of the scanning observations. But it is worth noting that this systematic error should be a conservative estimate due to the possible fluctuations of Crab. Figure \ref{fig:sensi} shows the relationships of the SNRs and fluxes of the three telescopes. It can be seen that the highest values of the SNRs are correlated with the source fluxes for all the three telescopes. Due to the various scan tracks, the exposure times of the scanned sources in the scanning observations are different, thus the SNRs of the scanned sources with the same fluxes can also be very different. From another point of view, there is also a flux range with a fixed SNR, and the lower limit of the flux range with SNR~=~5 can be considered as a rough estimation of the sensitivity at the 5$\sigma$ level. As shown in Figure \ref{fig:sensi}, the 5$\sigma$ sensitivities for an individual scanning observation can be estimated as $\sim$~3~mCrab, $\sim$~20~mCrab and $\sim$~18~mCrab for LE, ME and HE, respectively. With the Crab spectral parameters in \citet{2017ApJ...841..56}, these 5$\sigma$ sensitivities correspond to $\rm \sim7.6\times10^{-11}~erg~cm^{-2}~s^{-1}$ ($1-6$ keV), $\rm \sim4.0\times10^{-10}~erg~cm^{-2}~s^{-1}$ ($7-40$ keV), $\rm \sim2.6\times10^{-10}~erg~cm^{-2}~s^{-1}$ ($25-100$ keV) for LE, ME and HE, respectively.

\begin{deluxetable*}{cccc}[htbp]
\tablecaption{Systematic errors and sensitivities in the Galactic plane scanning survey\label{tab:sensi}}
\tablecolumns{5}
\tablenum{4}
\tablewidth{0pt}
\tablehead{
\colhead{ } &
\colhead{LE} &
\colhead{ME} &
\colhead{HE}
}
\startdata
Energy range (keV) & $1-6$ & $7-40$ & $25-100$ \\
Systematic error\tablenotemark{a} ($\%$) & 1.8 & 1.6 & 2.7 \\
Sensitivity\tablenotemark{b} (mCrab) & $\sim3$ & $\sim20$ & $\sim18$ \\
Sensitivity\tablenotemark{b} $(\rm 10^{-10}~erg~cm^{-2}~s^{-1})$ & $\sim0.76$ & $\sim4.0$ & $\sim2.7$ \\
\enddata
\tablenotetext{a}{For the long-term light curve monitoring.}
\tablenotetext{b}{For an individual scanning observation with $\emph{v} = 0^{\circ}.06~s^{-1}$, $\rm \emph{d} = 0^{\circ}.4$ and $\rm \emph{R} = 7^{\circ}$ at the 5$\sigma$ level.}
\end{deluxetable*}


\begin{figure*}[htbp]
\gridline{\fig{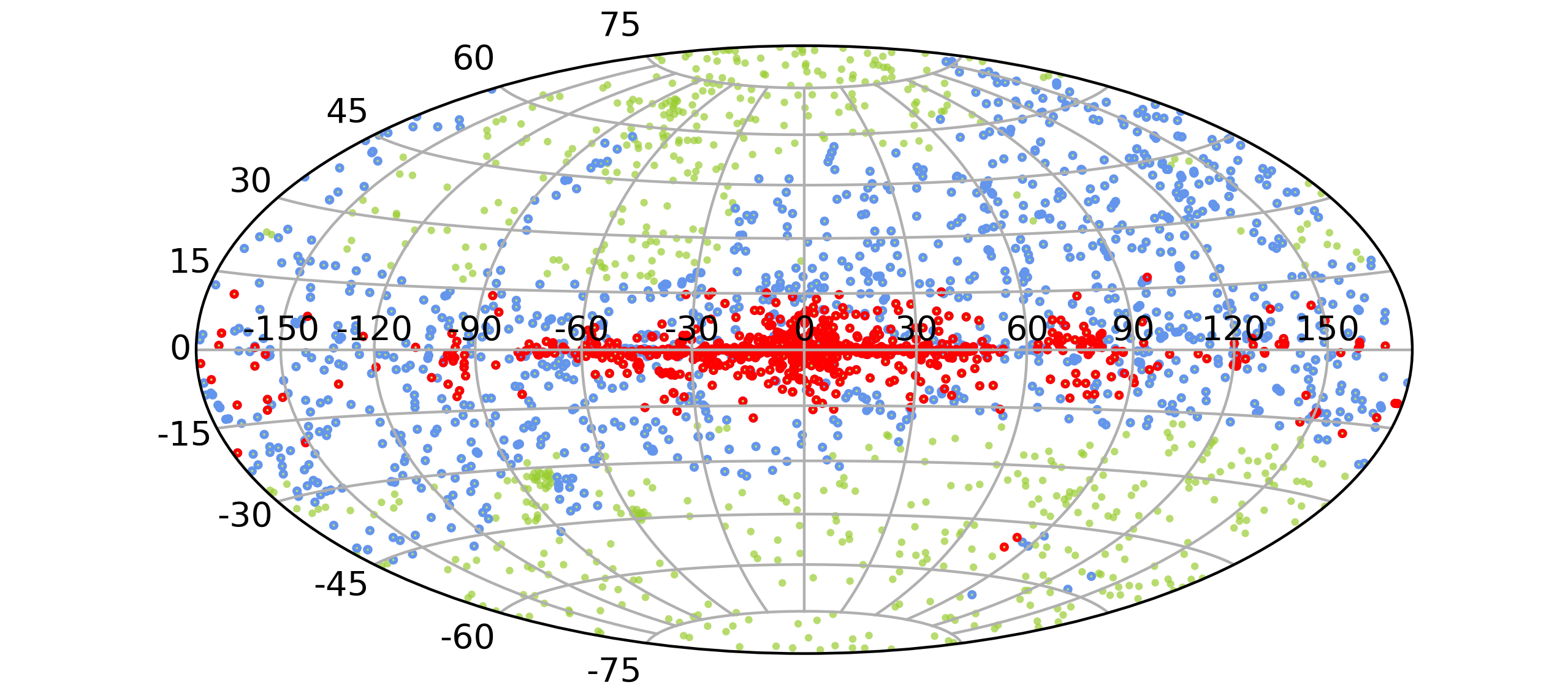}{0.7\textwidth}{(a) LE}}
\gridline{\fig{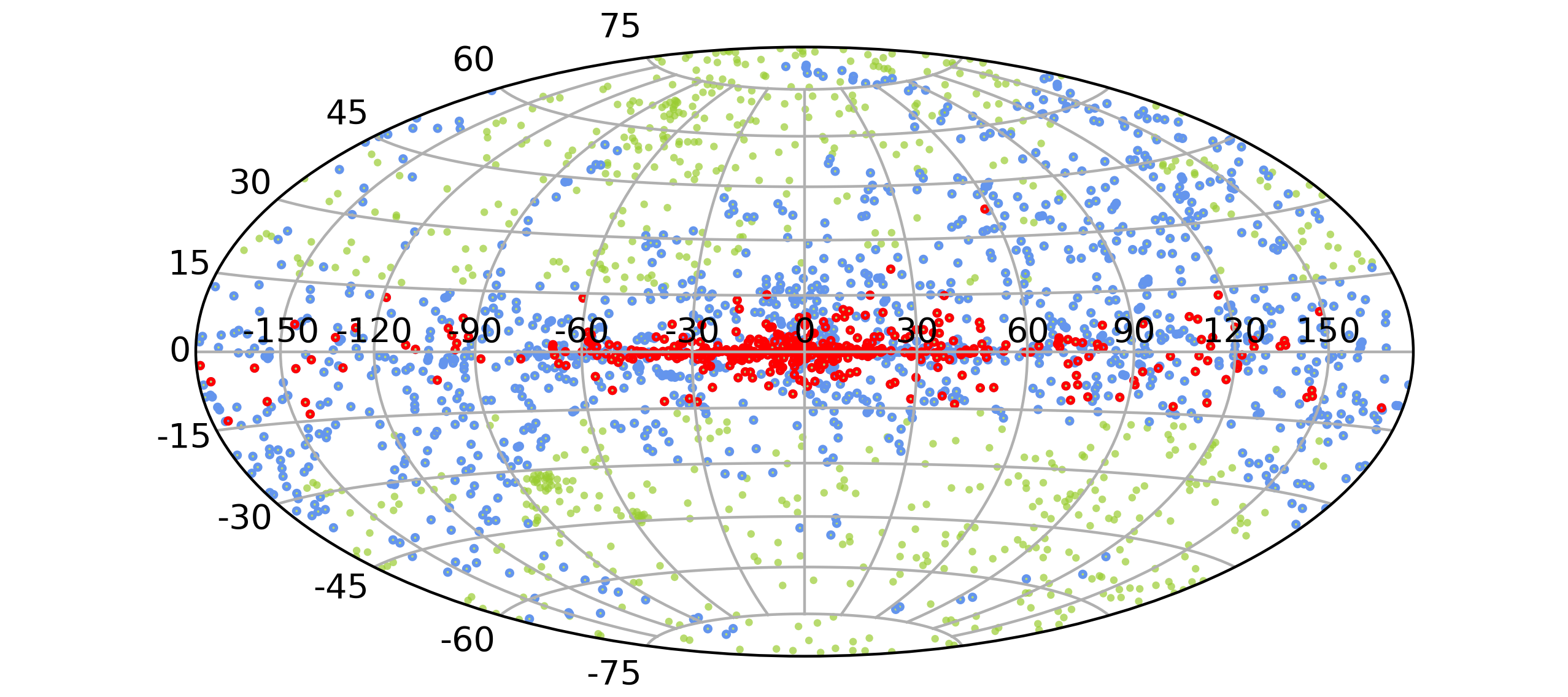}{0.7\textwidth}{(b) ME}}
\gridline{\fig{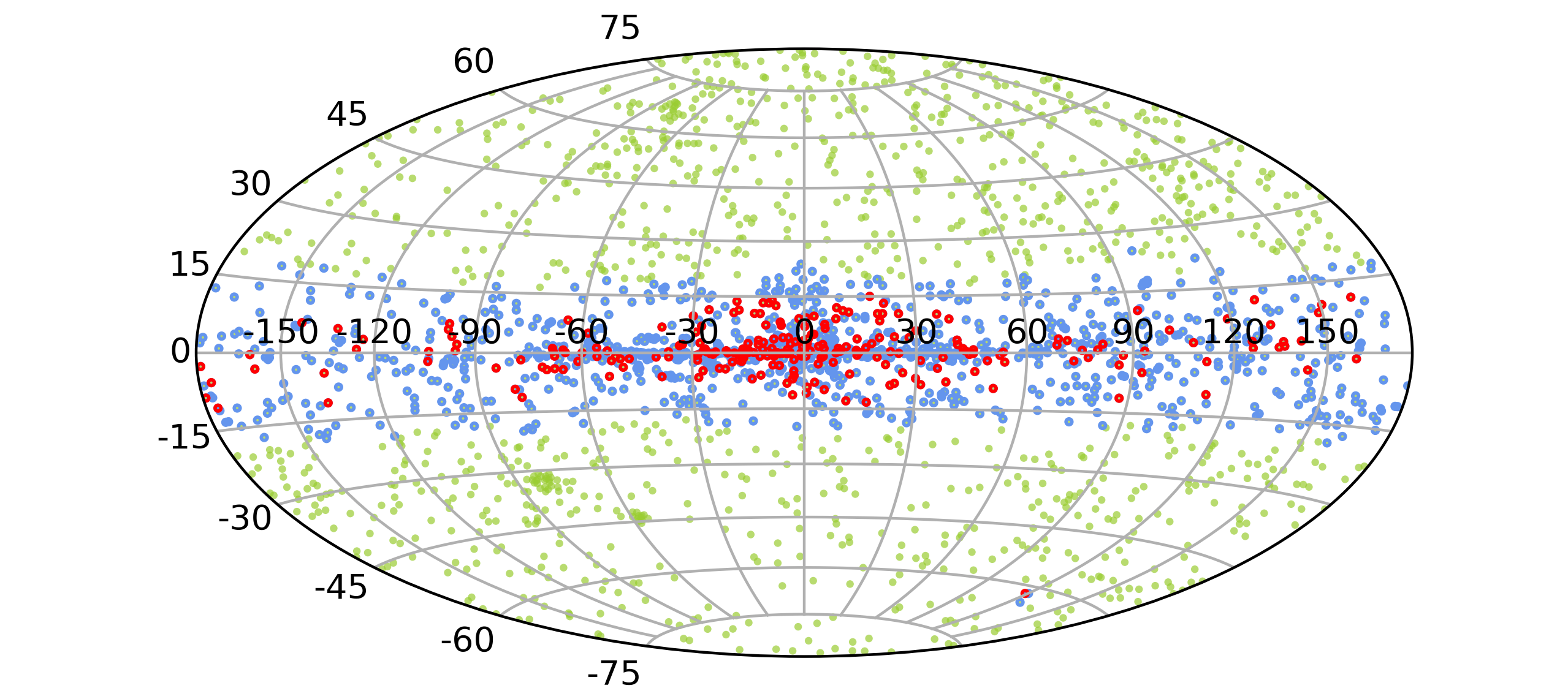}{0.7\textwidth}{(c) HE}}
\caption{Source positions in the Galactic coordinate system. The red and blue points refer to the strong (average $\rm SNR > 3$) and weak (average $\rm SNR < 3$) sources monitored by \emph{Insight-HXMT}, respectively; the green points refer to the sources in \emph{Swift}, \emph{INTEGRAL} and MAXI catalog.  \label{fig:smap}}
\end{figure*}

Among the data that has been processed, the number of the strong sources (average ${\rm SNR}>3$) $/$ weak sources (average ${\rm SNR}<3$) are $840/1231$, $410/1158$, and $246/985$ for LE, ME and HE, respectively. The positions of the sources monitored by LE, ME and HE are shown in Figure \ref{fig:smap}. It worth noting that the sources out of the Galactic plane are also monitored as \emph{Insight-HXMT} is in slew phase (attitude adjustment) before/after the scanning observation.

\section{Conclusions} \label{sec:con}
The Galactic plane scanning survey is one of the most important tasks of \emph{Insight-HXMT}. During the two-year operation in orbit of \emph{Insight-HXMT}, more than 1000 scanning observations have been performed to monitor the known sources and to search for new X-ray transients. A complete process has been developed to analysis the scanning data that mainly contains the data reduction and the PSF fitting. In addition to the standard data reduction of \emph{Insight-HXMT}, several essential operations are added to obtain more reliable data of the scanning observations, such as the removing of particle events for LE and ME in GTI selection, as well as the dynamic good pixel selection for ME scanning data. The fluxes and locations of the sources in a scanned region can be obtained by fitting the light curves with the PSF model that needs to be calibrated continuously. Up to September 2019, we have monitored more than 800 known sources and obtained their long-term light curves. Several new X-ray source candidates are also discovered, some of which can be the X-ray transients with very short time scale, which need to be further confirmed and thus not reported here.

\acknowledgments
This work made use of the data from the \emph{Insight-HXMT} mission, a project funded by China National Space Administration (CNSA) and the Chinese Academy of Sciences (CAS). The authors thank supports from the National Program on Key Research and Development Project (Grant No. 2016YFA0400802) and the National Natural Science Foundation of China under Grants No. U1838202, U1838201 and 11703028.


\newpage

\begin{thebibliography}{}
\bibitem[Baumgartner et al. (2013)]{2013ApJS...207...19} Baumgartner, W. H., Tueller, J., Markwardt, C. B., et al.,\ 2013. The 70 Month Swift-BAT All-sky Hard X-Ray Survey. The Astrophysical Journal Supplement, 207(2), 19. DOI: 10.1088/0067-0049/207/2/19.
\bibitem[Bird et al. (2016)]{2016ApJS...223...15} Bird, A. J., Bazzano, A., Malizia, A., et al.,\ 2016. The IBIS Soft Gamma-Ray Sky after 1000 Integral Orbits. The Astrophysical Journal Supplement Series, 223(1), 15. DOI: 10.3847/0067-0049/223/1/15.
\bibitem[Cao et al. (2020)]{2019Cao} Cao, X. L., Jiang, W. C., Meng, B., et al.,\ 2020. The Medium Energy (ME) X-ray telescope onboard the \emph{Insight-HXMT} astronomy satellite. Sci China-Phys Mech Astron, 63, 4: 249504. arXiv:1910.04451. DOI: 10.1007/s11433-019-1506-1.
\bibitem[Chen et al. (2020)]{2019Chen} Chen, Y., Cui, W. W., Li, W., et al.,\ 2020. The Low Energy X-ray telescope (LE) onboard the \emph{Insight-HXMT} astronomy satellite. Sci China-Phys Mech Astron, 63, 4: 249505. arXiv:1910.08319. DOI: 10.1007/s11433-019-1469-5.
\bibitem[HXMT User Analysis Software Group (2019)]{2019...2.01} HXMT User Analysis Software Group,\ 2019. The HXMT Data Reduction Guide v2.01. \url{http://www.hxmt.org/upl/doc/HXMT_data_reduction_guide_2_01.pdf}
\bibitem[Kouroubatzakis et al. (2017)]{2017ATel...10822} Kouroubatzakis, K.,Reig, P., Andrews, J., Zezas, A.,\ 2017.The optical counterpart to the new accreting pulsar Swift J0243.6+6124 is a Be star. ATel, 10822.
\bibitem[Li et al. (2009)]{2009CAA...33...333} Li, G., Wu, M., Zhang, S., Jin, Y.K.,\ 2009. Calculation for the Space Environment Background of HXMT. Chinese Astronomy and Astrophysics, 33(3), 333-346. DOI: 10.1016/j.chinastron.2009.07.013.
\bibitem[Liao et al. (2020)]{2019Liao} Liao, J. Y., Zhang, S., Lu, X. F., et al.,\ 2020. Background model for the high-energy telescope of \emph{Insight-HXMT}. JHEAp, submitted.
\bibitem[Liu et al. (2020)]{2019Liu} Liu, C. Z., Zhang, Y. F., Li, X. F., et al.,\ 2020. The High Energy X-ray telescope (HE) onboard the \emph{Insight-HXMT} astronomy satellite. Sci China-Phys Mech Astron, 63, 4: 249503. arXiv:1910.04955. DOI: 10.1007/s11433-019-1486-x.
\bibitem[Madsen et al. (2017)]{2017ApJ...841..56} Madsen, K. K., Forster, K., Grefenstette, B. W., et al.,\ 2017. The Astrophysical Journal, 841(1), 56. DOI: 10.1088/0067-0049/220/1/8.
\bibitem[Nang et al. (2020)]{2019N} Nang, Y., Liao, J. Y., Sai, N., et al.,\ 2020. In-orbit Calibration to the Point-Spread Function of \emph{Insight-HXMT}. JHEAp, in press. arXiv:2002.01097. DOI: 10.1016/j.jheap.2020.01.002
\bibitem[Ryan et al. (1988)]{1988PRB...34...396} Ryan, C., Clayton, E., Griffin, W., et al.,\ 1988. SNIP, a statistics-sensitive background treatment for the quantitative analysis of PIXE spectra in geoscience applications. Nuclear Instruments and Methods in Physics Research Section B, 34(3), 396-402. DOI: 10.1016/0168-583X(88)90063-8.
\bibitem[Xie et al. (2015)]{2015ASS...360...13} Xie, F., Zhang, J., Song, L.M., et al.,\ 2015. Simulation of the in-sight background for HXMT/HE. Astrophysics and Space Science, 360, 13.  DOI: 10.1007/s10509-015-2559-1.
\bibitem[Zhang et al. (2020)]{2019Zhang} Zhang, S. N., Li, T. P., Lu, F. J., et al.,\ 2020. Overview to \emph{the Hard X-ray Modulation Telescope} (\emph{Insight-HXMT}) Satellite. Sci China-Phys Mech Astron, 63, 4: 249502. DOI: 10.1007/s11433-019-1432-6.
\bibitem[Zhang et al. (2019)]{2019ApJ...879...61} Zhang, Y., Ge, M. Y., Song, L. M., et al.,\ 2019. \emph{Insight-HXMT} Observations of Swift J0243.6+6124 during Its 2017–2018 Outburst. The Astrophysical Journal, 879(1), 61. DOI: 10.3847/1538-4357/ab22b1
\bibitem[Zombeck (2006)]{2006CUP}Zombeck, M. V.,\ 2006. Handbook of space astronomy and astrophysics: Third Edition. Cambridge University Press.


\end{thebibliography}
\end{document}